\documentstyle[amsfonts]{article}
\def\cal{\frak }
\begin{document}
\title{Quantum algorithmic information theory}
\author{K. Svozil\\
 {\small Institut f\"ur Theoretische Physik}  \\
  {\small University of Technology Vienna }     \\
  {\small Wiedner Hauptstra\ss e 8-10/136}    \\
  {\small A-1040 Vienna, Austria   }            \\
  {\small e-mail: svozil@tph.tuwien.ac.at}\\
  {\small www: http://tph.tuwien.ac.at/$\widetilde{\;}$svozil}}
\maketitle

\begin{flushright}
{\scriptsize qait.tex}
\end{flushright}

\begin{abstract}
{\em
The agenda of quantum
algorithmic information theory, ordered `top-down,'
is the quantum halting amplitude, followed by the
quantum algorithmic information content, which in
turn requires the theory of quantum computation. The fundamental atoms
processed by quantum computation are the quantum bits
which are dealt with in
 quantum information theory.
The theory of quantum computation will be based upon a model
of universal quantum computer whose elementary unit is a two-port
interferometer capable of arbitrary $U(2)$ transformations.
Basic to all these considerations is quantum theory, in particular
Hilbert space quantum mechanics.
}
\end{abstract}

\section{Information is physical, so is computation}

The reasoning in constructive mathematics
\cite{bi-bi,bridges,bridges-94} and recursion theory,
at least insofar as their applicability to worldly things is concerned,
makes implicit assumptions about the operationalizability
of the entities of
discourse.
It is this postulated correspondence
between practical and theoretical objects, subsumed by
the Church-Turing thesis,
which confers power to
the formal methods.
Therefore, any finding in physics concerns the formal sciences; at least
insofar as they claim to be applicable in the physical
universe.
In this sense one might quite justifyably say that the Church-Turing
thesis is under permanent physical attack.\footnote{
 For an early discussion of this topic, see
Davis
 (cf. \cite{davis-58}, p. 11):
 \begin{quote}
 {\em `` $\ldots$ how can we ever exclude the possibility of our
 presented,
 some day (perhaps by some extraterrestrial visitors), with a (perhaps
 extremely complex) device or ``oracle'' that ``computes'' a
 non computable function?''
 }
 \end{quote}
A main theme of Landauer's  work has been the connections
between physics and computation; see, for example, his 1967 article
\cite{landauer-67}
{\em ``Wanted: a physically possible theory of physics,''} or his more
recent survey \cite{landauer} {\em ``Information is physical.''}
 See also  Rosen \cite{rosen}.
 As Deutsch puts it more recently (cf. \cite{deutsch-85}, p. 101),
 \begin{quote}
 {\em
 ``The reason why we find it possible to construct, say, electronic
 calculators, and indeed why we can perform mental arithmetic, cannot
 be found in mathematics or logic. {\em
 The reason is that the laws of physics `happen to' permit the
 existence of physical models for the operations of arithmetic}
 such as addition, subtraction and multiplication.
 If they did not, these familiar operations would be
 non-computable functions. We might still
 know {\em of} them and invoke them in mathematical proofs
 (which would presumably be called `non constructive') but we could
 not perform them.''
 }
 \end{quote}
}
Converseley, any feature of the (constructive or non-constructive
\cite{svozil-set}) formalism should correspond to some physically
operationalizable \cite{bridgman} property.

Hence, any theory of information, if applicable, has to deal with
entities
which are operational
\cite{bridgman,landauer,landauer-67,landauer-87,landauer-95}.
In Bridgman's words (cf. \cite{bridgman-reflections}, p. V),
 \begin{quote}
{\em
``the meaning of one's terms are to be found
by an analysis of the operations which one performs in applying the
term in concrete situations or in verifying the truth of statements or
in finding the answers to questions.''}
 \end{quote}
In particular, the fundamental atom  of information, the bit, must be
represented by whatever physical theories are available and must be
experimentally producible and manipulable by whatever physical
operations are available.

The classical digital computer, at least up to finite resources,
seems to be a canonical example for physical information representation
and processing.
Classical digital computers, however, are designed to behave
classically. That
is, if functioning correctly, certain of their physical states can be
mapped one-to-one onto the set of classical bit states. (This is achieved
by appropriately filtering out noise.) The set of
instructions implement the classical propositional calculus and so on.

In miniaturizing components, however, one encounters limits to the
quasi-classical domain. The alternative is either to stop
miniaturization before quantum effects become dominant, or to take the
quantum domain seriously. The latter
alternative (at least to the author) seems the only progressive one, but
it results in a head-on
collision with long-held classical properties.
Several long-held assumptions on the character of
information  have to be adapted. Furthermore,
the formal computational techniques in manipulating information have to
be revised.

This can be rather negatively perceived as a failure of the old models;
but I think that we are justified to think of it in very posive terms:
Physics, in particular quantum physics, stimulates us to re-consider our
conceptions. We could hope that the outcome will be new
tools and technologies in computing.

Indeed, right now, we are experiencing an attack on the ``Cook-Karp
thesis,'' putting into question the
robustness of the notion of tractability or polynomial time complexity
class
with respect to variations of ``reasonable'' models of computation.
In particular, factoring may require polynomial time on quantum
computers within
``reasonable statistics'' \cite{shor:94}.
 I would suspect that it is wise of mathematiciens
 and
computer scientists to keep an eye on new developments in physics, just
as we physicists are required to be open for the great advances in the
formal sciences.

\section{Hilbert space quantum mechanics}
``Quantization''
has been
introduced by
Max Planck
in 1900 \cite{planck:1901}.
Planck assumed
a {\em discretization} of the total energy
$U_N$ of
$N$
linear oscillators (``Resonatoren''),
$
U_N= P\epsilon \in \{ 0,\epsilon ,2\epsilon ,3\epsilon ,4\epsilon
,\ldots
\}
$,
where $P\in {\Bbb N}_0$ is zero or a positive
integer and $\epsilon$ stands for the {\em smallest quantum of energy}.
$\epsilon$ is a linear function of frequency $\nu$ and proportional to
Planck's fundamental constant $h$; i.e.,
$
\epsilon = h\nu
$.

In extension of Planck's discretized resonator energy model,
Einstein proposed a quantization of the
electromagnetic field. Every field mode of frequency $\nu$ could
carry a discrete  number of light quanta of energy
$h\nu$ per quantum.

The present quantum theory is still a continuum theory in many
respects: for infinite systems,
there is a continuity of field modes of frequency $\omega$.
Also the quantum theoretical coefficients characterizing the mixture
between orthogonal states, as well as
space and time and other coordinates
remain continuous --- all but one: action.
Thus, in the old days, discretization of
phase space
appeared to be a promising starting point for quantization.
In a 1916 article on the structure of physical phase space,
Planck emphasized that the quantum hypothesis should
not be interpreted at the level of energy quanta but at the level of
action quanta, according to the fact that the volume of $2f$-dimensional
phase space ($f$ degrees of freedom) is a positive integer of $h^f$
(cf. \cite{planck:1916}, p. 387),\footnote{
Again it is confirmed that the quantum hypothesis is not based on energy
elements but on action elements, according to the fact that the volume
of phase space has the dimension $h^f$.
}
\begin{quote}
Es best\"atigt sich auch hier wieder, da\ss~ die Quantenhypothese nicht
auf Energieelemente, sondern auf Wirkungselemente zu gr\"unden ist,
entsprechend dem Umstand, da\ss~ das Volumen des Phasenraumes die
Dimension von $h^f$ besitzt.
\end{quote}

The following is a very brief introduction to quantum mechanics for
logicians and computer scientists.\footnote{
Introductions to quantum mechanics can be found in
Feynman, Leighton \& M. Sands \cite{feynman-III},
Harris \cite{har},
Lipkin \cite{lipkin},
Ballentine  \cite{ba-89},
Messiah  \cite{messiah-61},
Dirac \cite{dirac},
Peres \cite{peres},
von Neumann \cite{v-neum-49}, and
Bell \cite{bell-87}, among many other expositions.
The history of quantum mechanics is reviewed by
Jammer \cite{jammer}.
Wheeler \& Zurek \cite{wheeler-Zurek:83} published a helpful resource
book.}
To avoid a shock from a too early exposition  to `exotic'
nomenclature prevalent in physics--the Dirac bra-ket notation--the
notation of Dunford-Schwartz
\cite{dunford-schwartz} is adopted.\footnote{
The bra-ket
notation introduced by Dirac which is widely used in physics. To
translate expressions into the bra-ket notation, the following
identifications have to be made: for the scalar product,
``$\langle  \equiv \;($'',
``$\rangle  \equiv \; )$'',
``$, \equiv \; \mid $''.
States are written as
$\mid  \psi \rangle  \equiv \psi$, operators as
$\langle  i\mid  A\mid  j \rangle  \equiv A_{ij}$.}

All quantum
mechanical entities are represented by objects
of Hilbert spaces \cite{v-neum-49}. A {\em Hilbert space} is a linear
vector space ${\cal H}$ over the field $\Phi$ of complex numbers
(with vector addition
and scalar multiplication), together  with a complex function
$(\cdot ,\cdot
)$, the {\em scalar} or {\em inner product}, defined on ${\cal
H}\times{\cal H}$ such that
(i)
$(x,x)=0$ if and only if $x=0$;
(ii)
$(x,x)\ge 0$ for all $x \in{\cal H}$;
(iii)
$(x+y,z)=(x,z)+(y,z)$ for all $x,y,z \in {\cal H}$;
(iv)
$(\alpha x,y)=\alpha (x,y)$ for all $x,y \in {\cal H}, \alpha \in \Phi$;
(v)
$(x,y)=\overline{(y,x)}$ for all $x,y \in {\cal H}$
($\overline{\alpha }$ stands for the complex conjugate of $\alpha$);
(vi)
If $x_n\in {\cal H}$, $n=1,2,\ldots$, and if $\lim_{n,m\rightarrow
\infty} (x_n-x_m,x_n-x_m)=0$, then there exists an $x\in {\cal H}$ with
$\lim_{n\rightarrow \infty} (x_n-x,x_n-x)=0$.

The following identifications between physical and theoretical objects
are made (a {\it caveat:} this is an incomplete list):

\begin{description}
\item[(I)]
 A {\em physical state} is represented by
a  vector of  the Hilbert space ${\cal H} $.
Therefore, if two vectors $x,y\in {\cal H}$ represent physical
states, their vector sum
$z=x+y\in{\cal H}$ represent a physical state as well.
This state $z$ is called the {\em coherent superposition} of state $x$
and
$y$. Coherent state superpositions will become most important in quantum
information theory.

\item[(II)]
{\em Observables} $A$ are represented by self-adjoint
operators $A$
on the Hilbert space ${\cal H}$ such that $(Ax,y)=(x,Ay)$ for all
$x,y\in {\cal H}$. (Observables and their corresponding operators are
identified.)

In what follows, unless stated differently, only
{\em finite} dimensional Hilbert spaces are considered.\footnote{
Infinite dimensional cases and continuous spectra are nontrivial
extensions of the finite
dimensional Hilbert space treatment. As a heuristic rule, it could be
stated that the sums become integrals, and the Kronecker delta function
$\delta_{ij}$
becomes the Dirac delta function $\delta (i-j)$, which is a
generalized function in the continuous variables $i,j$.
In the Dirac bra-ket notation, unity is given by
${\bf 1}=\int_{-\infty}^{+\infty} \vert i\rangle \langle i\vert \, di$.}
 Then, the vectors
corresponding to states can be written as usual vectors in complex
Hilbert space.
Furthermore, bounded
self-adjoint operators are  equivalent to bounded Hermitean operators.
They can be represented by matrices, and the self-adjoint
conjugation
is just transposition and complex conjugation of the matrix elements.

Elements $b_i,b_j\in {\cal H}$ of the set of orthonormal base vectors
satisfy
$(b_i, b_j) =\delta_{ij}$,
where $\delta_{ij}$ is the Kronecker delta function.
Any state $x$ can be written as a linear
combination of
the set of orthonormal base vectors $\{b_1,b_2,\cdots \}$,
i.e.,
$x =\sum_{i=1}^N   \beta_i b_i$, where $N$ is the dimension of ${\cal
H}$ and
$\beta_i=(b_i,x) \in \Phi$.
In the Dirac bra-ket notation, unity is given by
${\bf 1}=\sum_{i=1}^N \vert b_i\rangle \langle b_i\vert $.
Furthermore,
any Hermitean operator has a spectral representation
$A=\sum_{i=1}^N \alpha_i P_i$,
where the $P_i$'s  are orthogonal projection operators onto the
orthonormal eigenvectors $a_i$ of $A$ (nondegenerate
case).

As  infinite dimensional examples, take the
position operator  ${\vec {\frak x}}={\vec x}=(x_1,x_2,x_3)$, and the
momentum operator
${\vec {\frak p}_x}=
{\hbar \over i} {\vec \nabla}=
{\hbar \over i} \left(
{\partial \over \partial x_1},
{\partial \over \partial x_2},
{\partial \over \partial x_3}\right)
$, where $\hbar ={h\over 2\pi
}$. The scalar product is given by
$(
{\vec x},
{\vec y})=\delta^3 ({\vec x}-{\vec y})=
\delta(x_1-y_1)
\delta(x_2-y_2)
\delta(x_3-y_3)$.
The non-relativistic energy operator (Hamiltonian) is
${H}={{{\vec {\frak p}}} {{\vec {\frak p}}}\over
2m}+{V}(x)=
-\,
{\hbar^2
\over
2m}\nabla^2+V(x)$.

Observables are said to be {\em compatible} if they can be defined
simultaneously with arbitrary accuracy; i.e., if they are
``independent.'' A criterion for compatibility is the {\em commutator.}
Two observables ${A},{B}$ are compatible, if their {\em
commutator} vanishes; i.e.,
if $\left[
{A},
{B}
\right] =
{A}
{B}  -
{B}
{A}   =0$.
For example, position and momentum operators\footnote{the expressions
should be interpreted in the sense of
operator equations; the operators themselves act on states.}
$
\left[
{{\frak x}},
{{\frak p_x}}
\right] =
{{\frak x}}
{{\frak p_x}}-
{{\frak p_x}}
{{\frak x}} =
x
{\hbar \over i} {\partial \over \partial x}-
{\hbar \over i} {\partial \over \partial x}
x
=i\, \hbar
\neq 0
$
and thus do not commute. Therefore, position and momentum of a state
cannot be measured simultaneously with arbitrary accuracy.
It can be shown that this property gives rise to the {\em Heisenberg
uncertainty relations}
$
\Delta x
\Delta p_x \ge {\hbar \over 2}
$,
where
$
\Delta x
$
and
$
\Delta p_x
$
is given by
$
\Delta x =\sqrt{\langle x^2\rangle -\langle x\rangle ^2}
$
and
$
\Delta p_x =\sqrt{\langle p_x^2\rangle -\langle p_x\rangle ^2}
$,  respectively.
The expectation value
 or average value
$\langle \cdot \rangle $
is
defined in
{\bf
(V)} below.

It has recently been demonstrated that
(by an analog embodiment using
particle beams) every self-adjoint operator in a finite dimensional Hilbert
space can be experimentally realized \cite{rzbb}.

\item[(III)]
The result of any single measurement of the observable $A$
on a state $x\in {\cal H}$
can only be one of the real eigenvalues of the corresponding
Hermitean operator $A$.
If $x$ is in a coherent superposition of eigenstates of $A$, the
particular outcome of any such single measurement is indeterministic;
i.e.,
it cannot be predicted with certainty. As a
result of the measurement,
the system is in the state which corresponds to the eigenvector $a_n$ of
$A$ with the associated real-valued eigenvalue
$\alpha_n$; i.e., $Ax=\alpha_n a_n$ (no summation convention here).

This ``transition'' $x\rightarrow a_n$ has given rise to speculations
concerning the
``collapse
of the wave function (state).''  But, as has been argued recently
(cf. \cite{greenberger2}),
 it is
possible to reconstruct coherence; i.e., to ``reverse the collapse of
the wave function (state)'' if the process of measurement is
reversible. After this reconstruction, no information about the
measurement must be left, not even in principle.
How did Schr\"odinger, the creator of wave mechanics, perceive the
$\psi$-function? In his
1935 paper
``Die Gegenw\"artige
Situation in der Quantenmechanik'' (``The present situation in quantum
mechanics''
\cite{schrodinger}, p. 53), Schr\"odinger states,\footnote{
{\em The $\psi$-function as expectation-catalog:}
$\ldots$
In it [[the $\psi$-function]] is embodied the momentarily-attained sum
of theoretically based future expectation, somewhat as laid down in a
{\em catalog.}
$\ldots$
For each measurement one is required to ascribe to the $\psi$-function
($=$the prediction catalog) a characteristic, quite sudden change,
which {\em depends on the measurement result obtained,} and so {\em
cannot be forseen;} from which alone it is already quite clear
that this second kind of change of the $\psi$-function has nothing
whatever in common with its orderly development {\em between} two
measurements. The abrupt change [[of the $\psi$-function ($=$the
prediction catalog)]] by measurement $\ldots$ is the most interesting
point of the entire theory. It is precisely {\em the} point that demands
the break with naive realism. For {\em this} reason one cannot put the
$\psi$-function directly in place of the model or of the physical thing.
And indeed not because one might never dare impute abrupt unforseen
changes to a physical thing or to a model, but because in the realism
point of view observation is a natural process like any other and cannot
{\em per se} bring about an interruption of the orderly flow of natural
events.
}
\begin{quote}
{\em Die $\psi$-Funktion als Katalog der Erwartung:}
$\ldots$
Sie [[die $\psi$-Funktion]] ist jetzt das Instrument zur Voraussage der
Wahrscheinlichkeit von Ma\ss zahlen. In ihr ist die jeweils erreichte
Summe theoretisch begr\"undeter Zukunftserwartung verk\"orpert,
gleichsam wie in einem {\em Katalog} niedergelegt.
$\ldots$
Bei jeder Messung ist man gen\"otigt, der $\psi$-Funktion ($=$dem
Voraussagenkatalog eine eigenartige, etwas pl\"otzliche Ver\"anderung
zuzuschreiben, die von der {\em gefundenen Ma\ss zahl} abh\"angt und
sich {\em nicht vorhersehen l\"a\ss t;} woraus allein schon deutlich
ist, da\ss~ diese zweite Art von Ver\"anderung der $\psi$-Funktion mit
ihrem regelm\"a\ss igen Abrollen {\em zwischen} zwei Messungen nicht das
mindeste zu tun hat. Die abrupte Ver\"anderung durch die Messung
$\ldots$ ist der interessanteste Punkt der ganzen Theorie. Es ist genau
{\em der} Punkt, der den Bruch mit dem naiven Realismus verlangt. Aus
{\em diesem} Grund kann man die $\psi$-Funktion {\em nicht} direkt an
die Stelle des Modells oder des Realdings setzen. Und zwar nicht etwa
weil man einem Realding oder einem Modell nicht abrupte unvorhergesehene
\"Anderungen zumuten d\"urfte, sondern weil vom realistischen Standpunkt
die Beobachtung ein Naturvorgang ist wie jeder andere und nicht per se
eine Unterbrechung des regelm\"a\ss igen Naturlaufs hervorrufen darf.
\end{quote}
It therefore seems not unreasonable to state that, epistemologically,
quantum mechanics is more a theory of knowledge of an
(intrinsic) observer rather than the platonistic physics ``God knows.''
The  wave function, i.e., the state of the physical system in a
particular
representation (base), is a representation of the observer's knowledge;
it is a representation or name or code or index of
the information or knowledge the observer has access to.

\item[(IV)]
The probability $P_y(x)$ to find a system represented by state $x$
in some state $y$ of an orthonormalized basis is given by
$P_y(x)=\vert (x,y) \vert^2 $.

\item[(V)]
The {\em average value} or {\em expectation value} of an observable
$A$ in the state
$ x$
is given by
$\langle A\rangle_ x =
\sum_{i=1}^N \alpha_i
\vert (x,a_i) \vert^2$.

\item[(VI)]
The dynamical law or equation of motion can be written in the form
$x (t) =Ux (t_0) $,
where $U^\dagger =U^{-1}$ (``$\dagger $ stands for transposition and
complex conjugation) is a
linear {\em unitary evolution operator}.

The {\em Schr\"odinger equation}
$
i\hbar {\partial \over \partial t}  \psi (t)    =
H \psi (t) $
 is obtained by identifying $U$ with
$U=e^{-iHt/\hbar }$,
where $H$ is a self-adjoint  Hamiltonian (``energy'') operator,
by differentiating the equation of motion
with respect to the time variable $t$;
i.e.,
$
 {\partial \over \partial t} \psi (t) =-\,{iH\over
\hbar
}e^{-i{H}t/\hbar}
\psi (t_0 ) = -\,{i{H}\over \hbar } \psi (t)
$.
In terms of the
set of orthonormal base vectors $\{ B-1, b_2, \ldots
\}$, the Schr\"odinger equation can be written as
$i\hbar {\partial \over \partial t} ( b_i , \psi (t) )   =
\sum_{j}
H_{ij}( b_j, \psi (t) )$.
In the case of position base states $\psi (x,t)=( x, \psi (t)
)$, the Schr\"odinger equation takes on the form
 $
i\hbar {\partial \over \partial t}  \psi (x,t) =
{H} \psi (x,t)=\left[{{{\frak p}} {{\frak p}}\over 2m}+
{V}(x)\right]\psi (x,t) = \left[-\,{\hbar^2
\over
2m}\nabla^2+V(x)\right] \psi (x,t)$.

For stationary $ \psi_n
(t)=
e^{-(i/\hbar )E_nt}  \psi_n $, the Schr\"odinger equation
can be brought into its time-independent form
$H\, \psi_n
=
E_n\, \psi_n $.
Here,
$i\hbar {\partial \over \partial t} \psi_n (t)
=
E_n \, \psi_n (t) $  has been used;
$E_n$
and $\psi_n $
stand for the $n$'th eigenvalue and eigenstate of
$H$, respectively.

Usually, a physical problem is defined by the Hamiltonian ${H}$.
The problem of finding the physically relevant states reduces to finding
a complete set of eigenvalues and eigenstates of ${H}$.
Most elegant solutions utilize the symmetries of the problem, i.e., of
${H}$. There exist two ``canonical'' examples, the $1/r$-potential
and
the harmonic oscillator potential, which can be solved wonderfully by
this methods (and they are presented over and over again in standard
courses of quantum mechanics), but not many more. (See \cite{davydov}
for a detailed treatment of various Hamiltonians ${H}$.)
\end{description}

For a quantum mechanical treatment of a two-state system, see appendix
\ref{tws}.
For a review of the quantum theory of multiple particles, see appendix
\ref{a:2}.

\section{Quantum information theory}
The fundamental atom of information is the quantum bit, henceforth
abbreviated by the term `qbit'. As we shall see, qbits feature quantum
mechanics `in a nutshell.'

Classical information theory (e.g., \cite{hamming}) is based on the
classical bit as
fundamental atom. This classical bit, henceforth called
{\em cbit,} is in one of two
classical states $t$ (often interpreted as ``true'') and $f$ (often
interpreted as ``false'').
It is customary to code the classical logical states by
$\ulcorner t\urcorner =1$ and
$\ulcorner f\urcorner =0$ ($
\ulcorner
s
\urcorner$ stands for the code of $s$).
The states can, for instance, be realized by some
condenser who is discharged ($\equiv$ cbit state $0$) or charged
($\equiv$ cbit state $1$).

In quantum information theory (cf.
\cite{a:8,deutsch-85,f-85,peres-85,b-86,m-86,deutsch-89,deutsch:92}),
the most elementary unit of information is
 the {\em quantum bit,}
henceforth called {\em qbit}.
Qbits can be physically represented by a coherent
superposition
of the two orthonormal\footnote{
$(t,t)=(f,f)=1$ and $(t,f)=0$.}
 states $t$ and $f$.
The qbit states
\begin{equation}
x_{\alpha ,\beta}  =\alpha t+\beta f
\end{equation}
form a continuum, with
$ \vert \alpha \vert^2+\vert \beta \vert^2=1$, $\alpha ,\beta \in {\Bbb
C}$.

\subsection{Coding}
 Cbits can then be coded by
\begin{equation}
\ulcorner
x_{\alpha ,\beta }\urcorner  =(\alpha ,\beta )=
e^{i\varphi } (\sin \omega  ,e^{i\delta } \cos \omega )\quad ,
\end{equation}
with
$\omega ,\varphi ,\delta \in {\Bbb R}$.
Qbits can be identified with cbits as follows
\begin{equation}
(a,0)\equiv 1
\mbox{ and }
(0,b)\equiv 0
\quad , \qquad
\vert a\vert,
\vert b\vert =1\quad ,
\end{equation}
where the complex numbers $a$ and $b$ are of modulus one.
The quantum mechanical  states associated with the classical states $0$
and $1$ are mutually orthogonal.

Notice that, provided that $\alpha,\beta \neq 0$, a
qbit is not in a pure classical state. Therefore,
any practical determination of the qbit $x_{\alpha ,\beta }$
amounts to a measurement of the state amplitude of $t$ or $f$.
 Any such {\em single} measurement will be
indeterministic (provided again that $\alpha,\beta \neq 0$). That is,
the outcome of a single measurement occurs unpredictably.
Yet, according to the rules of quantum mechanics, the probabilities
that the qbit $x_{\alpha ,\beta }$ is measured in states $t$
and
$f$ is
$P_t(x_{\alpha ,\beta })=
\vert (x_{\alpha ,\beta },t)\vert^2
$ and
$P_f(x_{\alpha ,\beta })=
\vert (x_{\alpha ,\beta },f)\vert^2
=1-P_t(_{\alpha ,\beta })
$, respectively.

The classical and the quantum mechanical
concept of information differ from each other in several aspects.
Intuitively and classically, a unit of information is context-free.
That is, it is independent of what other information is or
might be present. A classical bit remains unchanged, no matter by what
methods it is inferred.
It obeys classical logic.
It can be copied. No doubts can be left.

By contrast,
quantum information is contextual \cite{kochen-specker}
A quantum bit may appear different,
depending on the method by which it is inferred.
Quantum bits cannot be copied or ``cloned''
\cite{wo-zu,dieks,mandel:83,mil-hard,glauber1,caves}.
Classical tautologies are not necessarily satisfied in quantum
information theory. Quantum bits obey quantum logic.
And, as has been argued before, they are coherent superpositions of
classical information.

\subsection{Reading the book of Nature---a short glance at the
prediction catalog}

To quote Landauer
\cite{landauer-89},
{\em ``What is measurement? If it is simply information transfer, that
is done all the time inside the computer, and can be done with arbitrary
little dissipation.''} And, one may add, {\em without destroying
coherence.}

Indeed, as has been shortly mentioned in {\bf (III)}, there is reason to
believe that---at least up to
a certain magnitude of complexity---any measurement can be ``undone'' by
a proper reconstruction of the wave-function. A necessary condition for
this to happen is that {\em all} information about the original
measurement is lost.
In Schr\"odinger's terms,
the prediction catalog
(the wave function) can be opened only at one particular page.
We may close
the prediction catalog
before reading this page. Then we can open
the prediction catalog
at another, complementary, page again.
By no way we can open
the prediction catalog at one page, read and (irreversibly) memorize the
page, close it; then open it at another, complementary, page.
(Two non-complementary pages which correspond to two co-measurable
observables can be read simultaneously.)

Can we then in some sense
``undo'' knowledge from conscious observation?
This question relates to a statement  by Wheeler (cf.
\cite{wheeler-Zurek:83}, p.
184)
that {\em ``no elementary phenomenon is a phenomenon until it is a[[n
irreversibly]] registered (observed) phenomenon.''}
Where does this irreversible observation take place? Since the physical
laws
 (with the possible exception of the weak force)
are time-reversible, the act of irreversible observation must, according
to Wigner \cite{wigner}, occur in the consciousness, thereby violating
quantum mechanics.

\section{Quantum recursion theory}

\subsection{Reversible computation and deletion of (q)bits}

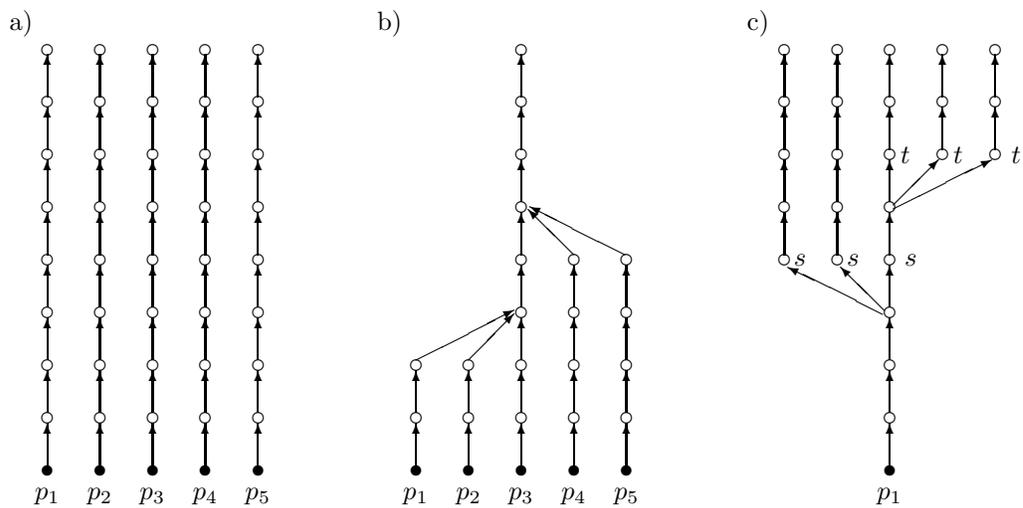
\begin{figure}
\begin{center}
\unitlength 0.70mm
\linethickness{0.4pt}
\begin{picture}(280.00,149.00)
\put(10.00,10.00){\circle*{2.00}}
\put(20.00,10.00){\circle*{2.00}}
\put(30.00,10.00){\circle*{2.00}}
\put(40.00,10.00){\circle*{2.00}}
\put(10.00,20.00){\circle{2.00}}
\put(10.00,11.00){\vector(0,1){8.00}}
\put(10.00,30.00){\circle{2.00}}
\put(10.00,21.00){\vector(0,1){8.00}}
\put(10.00,40.00){\circle{2.00}}
\put(10.00,31.00){\vector(0,1){8.00}}
\put(10.00,50.00){\circle{2.00}}
\put(10.00,41.00){\vector(0,1){8.00}}
\put(10.00,60.00){\circle{2.00}}
\put(10.00,51.00){\vector(0,1){8.00}}
\put(10.00,70.00){\circle{2.00}}
\put(10.00,61.00){\vector(0,1){8.00}}
\put(10.00,80.00){\circle{2.00}}
\put(10.00,71.00){\vector(0,1){8.00}}
\put(10.00,90.00){\circle{2.00}}
\put(10.00,81.00){\vector(0,1){8.00}}
\put(20.00,20.00){\circle{2.00}}
\put(20.00,11.00){\vector(0,1){8.00}}
\put(20.00,30.00){\circle{2.00}}
\put(20.00,21.00){\vector(0,1){8.00}}
\put(20.00,40.00){\circle{2.00}}
\put(20.00,31.00){\vector(0,1){8.00}}
\put(20.00,50.00){\circle{2.00}}
\put(20.00,41.00){\vector(0,1){8.00}}
\put(20.00,60.00){\circle{2.00}}
\put(20.00,51.00){\vector(0,1){8.00}}
\put(20.00,70.00){\circle{2.00}}
\put(20.00,61.00){\vector(0,1){8.00}}
\put(20.00,80.00){\circle{2.00}}
\put(20.00,71.00){\vector(0,1){8.00}}
\put(20.00,90.00){\circle{2.00}}
\put(20.00,81.00){\vector(0,1){8.00}}
\put(30.00,20.00){\circle{2.00}}
\put(30.00,11.00){\vector(0,1){8.00}}
\put(30.00,30.00){\circle{2.00}}
\put(30.00,21.00){\vector(0,1){8.00}}
\put(30.00,40.00){\circle{2.00}}
\put(30.00,31.00){\vector(0,1){8.00}}
\put(30.00,50.00){\circle{2.00}}
\put(30.00,41.00){\vector(0,1){8.00}}
\put(30.00,60.00){\circle{2.00}}
\put(30.00,51.00){\vector(0,1){8.00}}
\put(30.00,70.00){\circle{2.00}}
\put(30.00,61.00){\vector(0,1){8.00}}
\put(30.00,80.00){\circle{2.00}}
\put(30.00,71.00){\vector(0,1){8.00}}
\put(30.00,90.00){\circle{2.00}}
\put(30.00,81.00){\vector(0,1){8.00}}
\put(40.00,20.00){\circle{2.00}}
\put(40.00,11.00){\vector(0,1){8.00}}
\put(40.00,30.00){\circle{2.00}}
\put(40.00,21.00){\vector(0,1){8.00}}
\put(40.00,40.00){\circle{2.00}}
\put(40.00,31.00){\vector(0,1){8.00}}
\put(40.00,50.00){\circle{2.00}}
\put(40.00,41.00){\vector(0,1){8.00}}
\put(40.00,60.00){\circle{2.00}}
\put(40.00,51.00){\vector(0,1){8.00}}
\put(40.00,70.00){\circle{2.00}}
\put(40.00,61.00){\vector(0,1){8.00}}
\put(40.00,80.00){\circle{2.00}}
\put(40.00,71.00){\vector(0,1){8.00}}
\put(40.00,90.00){\circle{2.00}}
\put(40.00,81.00){\vector(0,1){8.00}}
\put(50.00,10.00){\circle*{2.00}}
\put(50.00,20.00){\circle{2.00}}
\put(50.00,11.00){\vector(0,1){8.00}}
\put(50.00,30.00){\circle{2.00}}
\put(50.00,21.00){\vector(0,1){8.00}}
\put(50.00,40.00){\circle{2.00}}
\put(50.00,31.00){\vector(0,1){8.00}}
\put(50.00,50.00){\circle{2.00}}
\put(50.00,41.00){\vector(0,1){8.00}}
\put(50.00,60.00){\circle{2.00}}
\put(50.00,51.00){\vector(0,1){8.00}}
\put(50.00,70.00){\circle{2.00}}
\put(50.00,61.00){\vector(0,1){8.00}}
\put(50.00,80.00){\circle{2.00}}
\put(50.00,71.00){\vector(0,1){8.00}}
\put(50.00,90.00){\circle{2.00}}
\put(50.00,81.00){\vector(0,1){8.00}}
\put(80.00,10.00){\circle*{2.00}}
\put(90.00,10.00){\circle*{2.00}}
\put(100.00,10.00){\circle*{2.00}}
\put(110.00,10.00){\circle*{2.00}}
\put(80.00,20.00){\circle{2.00}}
\put(80.00,11.00){\vector(0,1){8.00}}
\put(80.00,30.00){\circle{2.00}}
\put(80.00,21.00){\vector(0,1){8.00}}
\put(90.00,20.00){\circle{2.00}}
\put(90.00,11.00){\vector(0,1){8.00}}
\put(90.00,30.00){\circle{2.00}}
\put(90.00,21.00){\vector(0,1){8.00}}
\put(100.00,20.00){\circle{2.00}}
\put(100.00,11.00){\vector(0,1){8.00}}
\put(100.00,30.00){\circle{2.00}}
\put(100.00,21.00){\vector(0,1){8.00}}
\put(100.00,40.00){\circle{2.00}}
\put(100.00,31.00){\vector(0,1){8.00}}
\put(100.00,50.00){\circle{2.00}}
\put(100.00,41.00){\vector(0,1){8.00}}
\put(100.00,60.00){\circle{2.00}}
\put(100.00,51.00){\vector(0,1){8.00}}
\put(100.00,70.00){\circle{2.00}}
\put(100.00,61.00){\vector(0,1){8.00}}
\put(100.00,80.00){\circle{2.00}}
\put(100.00,71.00){\vector(0,1){8.00}}
\put(100.00,90.00){\circle{2.00}}
\put(100.00,81.00){\vector(0,1){8.00}}
\put(110.00,20.00){\circle{2.00}}
\put(110.00,11.00){\vector(0,1){8.00}}
\put(110.00,30.00){\circle{2.00}}
\put(110.00,21.00){\vector(0,1){8.00}}
\put(110.00,40.00){\circle{2.00}}
\put(110.00,31.00){\vector(0,1){8.00}}
\put(110.00,50.00){\circle{2.00}}
\put(110.00,41.00){\vector(0,1){8.00}}
\put(120.00,10.00){\circle*{2.00}}
\put(120.00,20.00){\circle{2.00}}
\put(120.00,11.00){\vector(0,1){8.00}}
\put(120.00,30.00){\circle{2.00}}
\put(120.00,21.00){\vector(0,1){8.00}}
\put(120.00,40.00){\circle{2.00}}
\put(120.00,31.00){\vector(0,1){8.00}}
\put(120.00,50.00){\circle{2.00}}
\put(120.00,41.00){\vector(0,1){8.00}}
\put(80.00,31.00){\vector(2,1){18.67}}
\put(90.00,31.00){\vector(1,1){8.67}}
\put(110.00,51.00){\vector(-1,1){8.67}}
\put(119.67,51.00){\vector(-2,1){18.00}}
\put(10.00,5.00){\makebox(0,0)[cc]{$p_1$}}
\put(20.00,5.00){\makebox(0,0)[cc]{$p_2$}}
\put(30.00,5.00){\makebox(0,0)[cc]{$p_3$}}
\put(40.00,5.00){\makebox(0,0)[cc]{$p_4$}}
\put(50.00,5.00){\makebox(0,0)[cc]{$p_5$}}
\put(80.00,5.00){\makebox(0,0)[cc]{$p_1$}}
\put(90.00,5.00){\makebox(0,0)[cc]{$p_2$}}
\put(100.00,5.00){\makebox(0,0)[cc]{$p_3$}}
\put(110.00,5.00){\makebox(0,0)[cc]{$p_4$}}
\put(120.00,5.00){\makebox(0,0)[cc]{$p_5$}}
\put(5.00,95.00){\makebox(0,0)[cc]{a)}}
\put(75.00,95.00){\makebox(0,0)[cc]{b)}}
\put(145.00,95.00){\makebox(0,0)[cc]{c)}}
\put(170.00,90.00){\circle{2.00}}
\put(170.00,80.00){\circle{2.00}}
\put(170.00,70.00){\circle{2.00}}
\put(170.00,60.00){\circle{2.00}}
\put(170.00,50.00){\circle{2.00}}
\put(170.00,40.00){\circle{2.00}}
\put(170.00,30.00){\circle{2.00}}
\put(170.00,20.00){\circle{2.00}}
\put(170.00,10.00){\circle*{2.00}}
\put(170.00,5.00){\makebox(0,0)[cc]{$p_1$}}
\put(170.00,11.00){\vector(0,1){8.00}}
\put(170.00,21.00){\vector(0,1){8.00}}
\put(170.00,31.00){\vector(0,1){8.00}}
\put(170.00,41.00){\vector(0,1){8.00}}
\put(170.00,51.00){\vector(0,1){8.00}}
\put(170.00,61.00){\vector(0,1){8.00}}
\put(170.00,71.00){\vector(0,1){8.00}}
\put(170.00,81.00){\vector(0,1){8.00}}
\put(280.00,131.00){\vector(0,1){8.00}}
\put(280.00,141.00){\vector(0,1){8.00}}
\put(180.00,70.00){\circle{2.00}}
\put(190.00,70.00){\circle{2.00}}
\put(180.00,80.00){\circle{2.00}}
\put(180.00,71.00){\vector(0,1){8.00}}
\put(180.00,90.00){\circle{2.00}}
\put(180.00,81.00){\vector(0,1){8.00}}
\put(190.00,80.00){\circle{2.00}}
\put(190.00,71.00){\vector(0,1){8.00}}
\put(190.00,90.00){\circle{2.00}}
\put(190.00,81.00){\vector(0,1){8.00}}
\put(150.00,50.00){\circle{2.00}}
\put(150.00,60.00){\circle{2.00}}
\put(150.00,51.00){\vector(0,1){8.00}}
\put(150.00,70.00){\circle{2.00}}
\put(150.00,61.00){\vector(0,1){8.00}}
\put(150.00,80.00){\circle{2.00}}
\put(150.00,71.00){\vector(0,1){8.00}}
\put(150.00,90.00){\circle{2.00}}
\put(150.00,81.00){\vector(0,1){8.00}}
\put(160.00,50.00){\circle{2.00}}
\put(160.00,60.00){\circle{2.00}}
\put(160.00,51.00){\vector(0,1){8.00}}
\put(160.00,70.00){\circle{2.00}}
\put(160.00,61.00){\vector(0,1){8.00}}
\put(160.00,80.00){\circle{2.00}}
\put(160.00,71.00){\vector(0,1){8.00}}
\put(160.00,90.00){\circle{2.00}}
\put(160.00,81.00){\vector(0,1){8.00}}
\put(168.67,39.67){\vector(-2,1){18.00}}
\put(169.00,40.33){\vector(-1,1){8.33}}
\put(170.67,60.33){\vector(1,1){8.67}}
\put(170.67,59.67){\vector(2,1){18.67}}
\put(174.00,50.00){\makebox(0,0)[cc]{$s$}}
\put(163.00,50.00){\makebox(0,0)[cc]{$s$}}
\put(153.00,50.00){\makebox(0,0)[cc]{$s$}}
\put(194.00,70.00){\makebox(0,0)[cc]{$t$}}
\put(183.00,70.00){\makebox(0,0)[cc]{$t$}}
\put(173.00,70.00){\makebox(0,0)[cc]{$t$}}
\end{picture}
\end{center}
\caption{The lowest ``root'' represents the
initial state interpretable as program. Forward computation represents
upwards motion
through a sequence of states represented by open circles. Different
symbols $p_i$ correspond to different initial states, that is, different
programs.
a) One-to-one computation.
b) Many-to-one junction which is information discarding. Several
computational paths, moving upwards, merge into one.
c) One-to-many computation is allowed only
 if no information is
created and discarded; e.g., in copy-type operations on blank memory.
\label{f-rev-comp}
}
\end{figure}

As a prelude to quantum computation, we briefly review classical
reversible computation
\cite{landauer-61,bennett-73,fred-tof-82,bennett-82,landauer-94}.
This type of computation is characterized by a single-valued inverse
transition function. That is, logical functions
are performed which
do not have a single-valued inverse, such as ${\tt AND}$ or ${\tt OR}$;
i.e., the input cannot be deduced from the output. Also deletion of
information or other many
(states)-to-one
(state) operations are irreversible.
Reversible calculation requires every single step to be reversible.
Figure \ref{f-rev-comp} (cf. \cite{landauer-94}) draws the difference
between one-to-one and many-to-one computation.
This logical irreversibility is associated with physical irreversibility
and requires a minimal heat generation of the computing machine.

It is possible to embed any irreversible computation in an appropriate
environment which makes it reversible. For instance, the computing agent
could keep the inputs of previous calculations in successive order.
It could save save all the information it would otherwise throw away.
 Or,
it could leave markers behind to identify its trail, the {\it H\"ansel
and Gretel} strategy described by Landauer \cite{landauer-94}. That, of
course, might amount to tremendous overhead in dynamical memory space
(and time) and would merely postpone the problem of throwing away
unwanted information. But, as was pointed out by Bennett
\cite{bennett-73}, this overhead could be circumvented by making the
computer to erase all intermediate results, leaving behind only the
desired output and the originally furnished input. Bennett's trick is to
do a computation reversibly, then copy its output\footnote{
Copying can be done
reversibly in classical physics, if the memory used for the copy is
initially blank. Quantum mechanically, this cannot be done on qbits;
cf below.}
and then, with one output as input for the reversible computation, run
the computation backwards. In order not to consume exceedingly large
intermediate storage resources, this strategy could be applied after
every single step. The price is a doubling of computation time, since it
requires one additional step for the back-computation.\footnote{
If an irreversible computing agent exists which computes the input from
a given output, then it is possible to translate an irreversible
computation from input to output into one which is reversible and erases
everything else except the final output, {\em including the
original input}; i.e., that simply maps inputs into outputs. For
details, see Bennett \cite{bennett-73,bennett-82}.}

\subsection{Selected features of quantum computation}

The following features are necessary but not sufficient qualities of
quantum computers.
\begin{description}
\item[(i)]
Input, output, program and memory are qbits;
\item[(ii)]
any computation (step) can be
represented by
a unitary transformation of the computer as a whole;
\item[(iii)]
because of the unitarity of the quantum evolution operator, any
computation is reversible.
Therefore, a deterministic computation can be performed by a quantum
computer if and only if it is reversible, i.e., if the program does
not involve ``deletion''of information
or  ``many-to-one'' operations
(cf. \cite{landauer}); only one-to-one operations are allowed;
\item[(iv)]
(in contradistinction to classical reversible computation)
unless classical, qbits cannot be copied; they are context-dependent
(cf. below);
\item[(v)]
measurements
may be carried out on any qbit at any stage of the computation.
But, unless classical, a qbit cannot be measured by a single
experiment with arbitrary accuracy (cf.
{\bf (III)} and {\bf (IV)}).
The computation process and the measurement have to be repeated in order
to obtain sufficient statistics.--Any such single measurement will yield
merely
a ``click'' on some counter, from which information about the qbit state
must
be inferred. Thereby, any single measurement is indeterminate, and
coherence is destroyed.
Therefore, it seems more proper to realize that there is no such
operational concept of ``a single
qbit.'' Because of complementarity, single qbits cannot be determined
precisely. What is henceforth called ``determination'' or
``measurement'' of a qbit is, in effect, the observation of a successive
number of such qbits, one after the other, from ``similar'' computation
processes (same preparation, same evolution). By performing these
measurements on
``similar'' qbits, one can
``determine'' this qbit within an
$\varepsilon$-neighborhood
only. The parameter $\varepsilon$ depends on the number of successive
measurements made;
\item[(vi)]
quantum parallelism: during a computation (step), a quantum computer
proceeds down all coherent paths at once;
\item[(vii)]
any subroutine must not leave around any qbits beyond it's computed
answer, because the computational paths with different residual
information can no longer interfere \cite{benn-94a}.
\end{description}

In order to appreciate  quantum
computation, one should make proper use of the latter
features--quantum parallelism, unerasability of
information, non-copying, context-dependence and impossibility to
directly measure the atoms of quantum information, the qbits, related to
quantum indeterminism.

Stated pointedly: the
quantum computation
``solution''
to a decision problem
may yield the classical bit values
at random.
It may depend on other qbits of information which are inferred.
It cannot be arbitrarily copied and, in this sense, is unique.

\subsubsection{Copying of quantum bits}

Can a non-classical qbit
be copied? No! ---
This answer amazes the classical
mind.\footnote{Copying of qbits would allow circumvention of the
Heisenberg uncertainty relation by measuring two incompatible
observables on two identical qbit copies. It would also allow
faster-than-light transmission of information, as pointed out by Herbert
\cite{herbert}. Herbert's suggestion stimulated the development of
``no-cloning theorems'' reviewed here.}
Informally speaking, the reason is that any attempt to copy a coherent
superposition
of states results either in a state reduction, destroying coherence,
or, most important of all, in the addition of noise which manifests
itself
as the spontaneous excitations of previously nonexisting field modes
\cite{wo-zu,dieks,mandel:83,mil-hard,glauber1,caves}.
Therefore, {\em qbits can be copied if and only if they
are (known to be) classical.
Only one-to-one computation processes depicted in Fig.
\ref{f-rev-comp}a) are allowed.}

This can be seen by a short calculation \cite{wo-zu} which requires
multi-quantum formalism developed in appendix \ref{a:2}.
A physical realization\footnote{the most elementary realization is a
one-mode field with the symbol ${\bf 0}$
corresponding to $\mid  0\rangle  $ (empty mode) and
${\bf 1}$ corresponding to $\mid  1\rangle  $ (one-quantum filled mode).}
 of the qbit state
is a two-mode boson field
with the identifications
\begin{eqnarray}
 x_{\alpha ,\beta}&=&\alpha t +\beta f
\quad ,\\
f&=& \mid  0_1,1_2\rangle
\quad ,\\
t &=& \mid  1_1,0_2\rangle
\quad .
\end{eqnarray}
The classical bit states are
$\vert 0_1,1_2\rangle $
 (field mode $1$ unfilled, field mode $2$
filled with one quantum)
and
$\vert 1_1,0_2\rangle       $
 (field mode $1$
filled with one quantum, field mode $2$ unfilled).

An ideal amplifier, denoted by $ A$, should be able to copy a
classical bit state; i.e., it should create an identical particle in the
same mode
\begin{equation}
A_i
\vert 0_1,1_2\rangle  \rightarrow
A_f
\vert 0_1,2_2\rangle
\quad ,
\qquad
A_i
\vert 1_1,0_2\rangle  \rightarrow
A_f
\vert 2_1,0_2\rangle
\quad .
\end{equation}
Here, $A_i$ and $A_f$ stand for the initial and the final state of the
amplifier.

What about copying a proper qbit; i.e., a {\em coherent superposition}
of the cbits
$f=\vert 0_1,1_2\rangle $ and
$t=\vert 1_1,0_2\rangle       $?
According to the quantum evolution law,
the corresponding amplification process should be
representable by a linear (unitary) operator; thus
\begin{equation}
A_i
(\alpha \vert 0_1,1_2\rangle  +
\beta \vert 1_1,0_2\rangle  )\rightarrow
 A_f
(a \vert 0_1,2_2\rangle   +
b \vert 2_1,0_2\rangle  )
\quad .
\label{l:ni1}
\end{equation}

Yet, the true copy of that qbit is the state
\begin{equation}
(\alpha \vert 0_1,1_2\rangle  +
\beta \vert 1_1,0_2\rangle  )^2 =
(\alpha \, a_2^\dagger +\beta \, a_1^\dagger )^2 \mid  0\rangle
=
\alpha^2 \vert 0_1,2_2\rangle   +
2\alpha \beta \vert 0_1,1_2\rangle
\vert 1_1,0_2\rangle       +
\beta^2 \vert 2_1,0_2\rangle
\quad .
\label{l:ni2}
\end{equation}
By comparing
(\ref{l:ni1}) with
(\ref{l:ni2}) it can be seen a reasonable (linear unitary
quantum mechanical evolution for an) amplifier which could copy a
qbit exists only if the qbit is classical.

A more detailed analysis
(cf. \cite{mandel:83,mil-hard}, in particular \cite{glauber1,caves})
reveals that the copying (amplification) process generates
an amplification of the signal but necessarily adds noise at the same
time.
This noise can be interpreted as spontaneous emission of field quanta
(photons) in the process of amplification.

One application of this feature is quantum cryptography
\cite{benn-82,benn-84,benn-92}. Thereby, the impossibility to copy qbits
is used for a cryptographic communication {\em via} quantum channels.

\subsubsection{Context dependence of qbits}
\label{q:context}
This section could be skipped at first reading.

Assume that in an EPR-type arrangement
\cite{epr}
one wants to measure  the product
$$
P=
m_x^1m_x^2
m_y^1m_y^2
m_z^1m_z^2
$$
of the direction of the spin components of each one of the two
associated particles
$1$ and $2$ along the $x$, $y$ and $z$-axes.
Assume that the operators are normalized such that $\vert m_i^j\vert=1$,
$i\in \{ x,y,z\}$, $j\in \{ 1,2\}$.
One way to determine $P$
is
measuring
and, based on these measurements,
``counterfactually inferring'' \cite{peres,mermin}
the three ``observables''
$
m_x^1m_y^2$, $
m_y^1m_x^2$ and $
m_z^1m_z^2$.
By multiplying them, one obtains $+1$.
Another, alternative, way to determine $P$
is
measuring
and, based on these measurements,
``counterfactually inferring''
the three ``observables''
$
m_x^1m_x^2$, $
m_y^1m_y^2$ and $
m_z^1m_z^2$.
By multiplying them, one obtains
$-1$.
In that way, one has obtained
either $P=1$ or $P=-1$.
Associate with $P=1$ the bit state zero ${\bf 0}$ and with $P=-1$ the
bit state ${\bf 1}$.
Then the bit is either in state zero or one, depending on the way
or {\em context} it was inferred.

This kind of contextuality is deeply rooted in the non-Boolean
algebraic structure of quantum propositions.
Note also that the above argument relies heavily on ``counterfactual
reasoning,'' because, for instance, only two of the six observables
$m_i^j$ can actually be experimentally determined.
Here, the term ``counterfactual reasoning'' \cite{peres,mermin} stands
for arguments involving results of {\em incompatible} experiments, i.e.,
experiments which could never be performed simultanuously, since the
associated operators do not commute. The results thus have to be {\em
inferred} rather than {\em measured}, and the existence of such
``elements of physical reality'' thus have to be tacitly assumed
\cite{epr}.

\subsection{Universal quantum computer based on the $U(2)$-gate}

The ``brute force'' method of obtaining a (universal) quantum
computer \cite{be,deutsch-85}
 by quantizing the ``hardware'' components  of a Turing machine
suffers from the same problem as its classical counterpart--it
seems technologically unreasonable to actually construct a universal
quantum device with
a ``scaled down'' (to nanometer size) model of a Turing machine in
mind.

We therefore pursue a more fundamental approach. Recall that an
arbitrary quantum time evolution in finite-dimensional Hilbert space
 is given by
$x (t) =Ux (t_0) $, where  $U$ is unitary.

It is well known that
any $n$-dimensional unitary matrix $U$ can be composed from elementary
unitary transformations in twodimensional subspaces of ${\Bbb C}^n$.
This is usually shown in the context of parameterization of the
$n$-dimensional unitary groups
(cf. \cite{murnaghan}, chapter 2 and
\cite{rzbb,reck-94}).
Thereby, a transformation in $n$-dimensional spaces is decomposed into
transformations in $2$-dimensional subspaces.
This amounts to a
successive array of $U(2)$ elements, which in their entirety forms an
arbitrary time evolution
$U(n)$ in n-dimensional Hilbert space.

It remains to be shown that the universal $U(2)$-gate is physically
operationalizable. This is done in appendix
\ref{a:u(2)}
in the framework of Mach-Zehnder interferometry.

The number of elementary $U(2)$-transformations is
polynomially bounded and does not exceed
$\left(
\begin{array}{c}
n\\2
\end{array}
\right) ={n\,(n-1)/ 2} =O(n^2)
$.

\subsection{Other models of universal quantum computation}

Deutsch \cite{deutsch-89} has proposed a model of universal computation
based on quantum computation networks.
In a recent paper, Barenco {\it et al.} \cite{barenco} show that
a set of gates that consists of all $U(2)$ (one-bit) quantum gates
and the two-bit exclusive-or gate (that maps Boolean values
$(x,y)$ to $(x,x \oplus y)$) is universal in the sense that all unitary
operations on arbitrarily many bits $n$ ($U(2^n$)) can be expressed as
compositions of these gates.

Thereby, the states
in a $2^n$-dimensional Hilbert space are constructed as the product
state of n particles in 2-dim Hilbert space, whereas the interferometric
approach using $U(2)$-gates introduced before is based on a single
particle state in $2^n$-dimensional Hilbert space.  In order to obtain
the mixing between different particle states,   xor-gates are needed.
The interferometric approach does not need xor-gates explicitly.

It has been claimed that certain $NP$-hard
problems such as factoring can be solved in polynomial time
\cite{shor:94}
on quantum computers. We shall not pursue these matters further
\cite{deutsch:92,d92a,be-va,be-br,be,cerni,shor:94}.

One of the most common models

\subsection{Nomenclature}
 Consider a (not necessarily universal) quantum computer $C$ and its
$i$th
 program $p_i$,
 which, at time $\tau \in {\bf Z}$, can be described by a quantum state
$C(\tau , p_i)$.
Let $C(p)=s$ stand for a computer $C$ with program $p$ which outputs $s$
in arbitrary long time.
In what follows we shall assume that the program $p_i$ is
coded {\em classically.} That is, we choose a finite code alphabet $A$
and denote by $A^\ast$ the set of all strings over $A$.
Any program $p_i$ is coded as a classical sequence
$
\ulcorner
p_i
\urcorner
=s_{1i}s_{2i}\cdots s_{ni}\in A^\ast $, $s_{ji}\in A$.
Whenever possible, $
\ulcorner
p_i
\urcorner
$ will be abbreviated by $p_i$.
We assume prefix coding
\cite{hamming,chaitin,svozil,calude}; i.e., the domain of $C$ is
prefix-free such that no admissible program is the prefix of another
admissible program.
Furthermore, without loss of generality, we consider only empty input
strings.
$\vert p\vert$ stands for the length of $p$.

\subsection{Diagonalization}

This is neither the place for a comprehensive review of the
diagonalization method (cf. \cite{rogers,odi}), nor suffices the
author's competence
for such an endeavor. Therefore, only a few hallmarks are stated.
As already G\"odel pointed out in his classical paper
on the incompleteness of arithmetic
\cite{godel1},
the undecidability theorems of formal logic \cite{davis-58}
(and the theory of recursive functions
\cite{rogers,odi})
are based on semantical paradoxes
such as the liar
\cite{bible} or Richard's paradox.
A proper translation of the semantic paradoxes results in
the diagonalization method. Diagonalization
has apparently first been applied by
Cantor to demonstrate the undenumerability of real numbers
\cite{cantor}. It has also been used by Turing for a proof
of the recursive undecidability of the halting problem \cite{turing}.

A brief review of the classical algorithmic argument will be given
first.  Consider a universal computer $C$. For the sake of
contradiction, consider an arbitrary algorithm
$B(X)$ whose input is a string of symbols $X$.  Assume that there exists
a ``halting algorithm'' ${\tt HALT}$ which is able to decide whether $B$
terminates on $X$ or not.
The domain of ${\tt HALT}$  is the set of legal programs.
The range of ${\tt HALT}$ are cbits (classical case) and qbits (quantum
mechanical case).

Using ${\tt HALT}(B(X))$ we shall construct another deterministic
computing agent $A$, which has as input any effective program $B$ and
which proceeds as follows:  Upon reading the program $B$ as input, $A$
makes a copy of it.  This can be readily achieved, since the program $B$
is presented to $A$ in some encoded form
$\ulcorner B\urcorner $,
i.e., as a string of
symbols.  In the next step, the agent uses the code
$\ulcorner B\urcorner $
 as input
string for $B$ itself; i.e., $A$ forms  $B(\ulcorner B\urcorner )$,
henceforth denoted by
$B(B)$.  The agent now hands $B(B)$ over to its subroutine ${\tt HALT}$.
Then, $A$ proceeds as follows:  if ${\tt HALT}(B(B))$ decides that
$B(B)$ halts, then the agent $A$ does not halt; this can for instance be
realized by an infinite {\tt DO}-loop; if ${\tt HALT}(B(B))$ decides
that $B(B)$ does {\em not} halt, then $A$ halts.

The agent $A$ will now be confronted with the following paradoxical
task:  take the own code as input and proceed.

\subsubsection{Classical case}
 Assume that $A$ is
restricted to classical bits of information.
To be more specific,
assume that ${\tt HALT}$ outputs the code of a cbit as follows
($\uparrow$ and $\downarrow$ stands for divergence and convergence,
respectively):
\begin{equation}
{\tt HALT} ( B(X) ) =\left\{
 \begin{array}{l}
0 \mbox{ if } B(X) \uparrow
\\
1 \mbox{ if } B(X) \downarrow \\
\end{array}
 \right.
\quad .
\label{el:halt}
\end{equation}

Then, whenever $A(A)$
halts, ${\tt HALT}(A(A))$ outputs $1$ and forces $A(A)$ not to halt.
Conversely,
whenever $A(A)$ does not halt, then ${\tt HALT}(A(A))$ outputs $0$
and steers
$A(A)$ into the halting mode.  In both cases one arrives at a complete
contradiction.  Classically, this contradiction can only be consistently
avoided by assuming the nonexistence of $A$ and, since the only
nontrivial feature of $A$ is the use of the peculiar halting algorithm
${\tt HALT}$, the impossibility of any such halting algorithm.

\subsubsection{Quantum mechanical case}

Recall that a quantum computer $C$ evolves according to a unitary
operator
$U$  such that ($\tau $ stands for the discrete time parameter)
$ C(\tau ,p_i)  =U C(\tau -1, p_i)
=U^t C(0,p_i)$.

As has been pointed out before, in quantum information theory
a qbit may be in a coherent
superposition
of the two classical states $t$ and $f$.
Due to this possibility of a coherent superposition of classical bit
states, the usual {\it reductio ad absurdum} argument breaks down.
Instead, diagonalization procedures in
quantum information theory yield qbit solutions which are fixed points
of the associated unitary operators.

In what follows it will be demonstrated how the task of the agent $A$
can be performed consistently if
$A$ is allowed to process quantum information.
To be more specific, assume that the output of the hypothetical
``halting algorithm'' is a halting qbit
\begin{equation}
{\tt HALT} ( B(X) ) = h_{\alpha , \beta}
\quad .
\end{equation}
One may think of   ${\tt HALT} ( B(X) )$ as a universal ``watchdog''
computer
$C'$
simulating $C$ and containing a dedicated {\em halting bit}, which it
 outputs
at every (discrete) time cycle
\cite{deutsch-85}.
 Alternatively, it can be
assumed that the computer $C$ contains its own halting bit
indicating whether it has completed its task or not.
Note that the halting qbit
$ h_{\alpha , \beta}
$ can be represented by a normalized\footnote{ $(
h_{\alpha , \beta},
h_{\alpha , \beta})=1$.}
 vector
 in twodimensional complex
Hilbert space spanned by the
 the orthonormal vectors
``$t $''
and
``$f $.''
Let the halting state $h_{ 1,0} =
t
$ (up to factors modulus  1) be the physical realization that the
computer has
``halted;''
likewise let
$h_{ 0,1} = f$ (up to factors modulus  1)
be the physical realization that the computer has not
``halted.''
Note that, since quantum computations are governed by unitary evolution
laws which are reversible, the halting state does not imply that the
computer does not change as time evolves. It just means that it has set
a signal --- the halting bit --- to indicated that it has finished its
task.
$\alpha $ and $\beta$ are complex numbers which are a quantum mechanical
measure
of the probability amplitude that the computer is in the halting and the
non-halting states, respectively.
The corresponding halting and non-halting probabilities are $\vert
a\vert^2$ and $\vert a\vert^2$, respectively.

Initially, i.e.,
at
$t=0$, the halting bit is prepared to be a 50:50 mixture of the
classical halting and non-halting states $t$ and $f$; i.e.,
$h_{1/\sqrt{2} , 1/\sqrt{2} }$. If later $C'$ finds that $C$ converges
(diverges) on $B(X)$, then the halting bit of $C'$ is set to the
classical value $t$ ($f$).

The emergence of fixed points can be demonstrated by a simple example.
Agent $A$'s diagonalization task can be formalized as
follows. Consider for the moment the action of diagonalization  on the
cbit states. (Since the qbit states are merely a coherent superposition
thereof, the action of diagonalization on qbits is straightforward.)
Diagonalization effectively transforms the cbit value $t$ into $f$ and
{\it vice versa.}
Recall that in equation
(\ref{el:halt}),  the state
$t$ has been identified
 with the halting state and the state $f$
with the non-halting
state. Since the halting state and the non-halting state exclude each
other,
$f,t$ can be identified with orthonormal basis vectors  in a
twodimensional vector space. Thus, the standard basis of
Cartesian coordinates can be chosen for a representation of $t$ and $f$;
i.e.,
\begin{equation}
t  \equiv
\left(
\begin{array}{c}
1 \\
0
 \end{array}
\right)
\mbox{ and }
f \equiv
\left(
\begin{array}{c}
0 \\
1
 \end{array}
\right) \quad .
\end{equation}

 The evolution representing diagonalization (effectively, agent
$A$'s task) can be expressed by the unitary operator $D$ by
\begin{equation}
D t  =  f \mbox{ and }
D f  =  t\quad .
\end{equation}
Thus, $D$ acts essentially as a ${\tt not}$-gate.
In the above state basis, $D$ can be represented as follows:
\begin{equation}
D=
\left(
\begin{array}{cc}
0 & 1\\
1 & 0
\end{array}
\right) \quad .
\end{equation}
$ D $ will be called {\em diagonalization} operator, despite the fact
that the only nonvanishing components are off-diagonal.

As has been pointed out earlier,
quantum information theory allows a coherent superposition
$ h_{\alpha ,\beta}  =\alpha t+\beta f $
of the
cbit states $t$ and $f$.
$D$ acts on cbits. It
has a
fixed point at the qbit state
\begin{equation}
h^\ast :=h_{ {1\over \sqrt{2}},{1\over \sqrt{2}} }  ={t+f\over \sqrt{2}}
\equiv
{1\over \sqrt{2}} \left(
\begin{array}{c}
1 \\
1
 \end{array}
\right) \quad .
\end{equation}
$h^\ast$
does not give rise to inconsistencies \cite{svozil-paradox}.
If agent $A$ hands over the fixed point state
$h ^\ast $ to the diagonalization
operator $D$, the same state
$h^\ast $ is recovered.
Stated differently, as long as the output of the ``halting
algorithm'' to input $A(A)$ is $h^\ast$, diagonalization does not
change it. Hence, even if the (classically) ``paradoxical'' construction
of diagonalization is maintained, quantum theory does not give rise to a
paradox, because the quantum range of solutions is larger than the
classical one.
Therefore,
standard proofs of the recursive unsolvability of the halting problem
do not apply if agent $A$ is allowed a qbit.

Another, less abstract, application for quantum information theory is
the handling of inconsistent information in databases.
Thereby,
two contradicting cbits of information
$t$ and
$f $ are resolved by the qbit
$h^\ast =
{(t+f)/ \sqrt{2}}$.
Throughout the rest of the computation the coherence is maintained.
After the processing, the result is obtained by an irreversible
measurement. The processing of qbits, however, would require an
exponential
space overhead on classical computers in cbit base \cite{feynman}.
Thus, in order to remain tractable,
the corresponding qbits should be implemented on
truly quantum universal computers.

It should be noted, however, that the fixed point qbit ``solution''
to the above halting problem,
as far as problem solving is concerned, is of not much practical help.
In particular, if one is interested in the ``classical'' answer whether
or not $A(A)$ halts,  then one ultimately has to perform an
irreversible measurement
on the fixed point state. This  causes a state reduction into the
classical states corresponding to $t$ and $f$.
Any single measurement will yield an indeterministic result.
There is a 50:50 chance that
the fixed point state will be either in $t$ or $f$, since
$P_t(
h ^\ast)=
P_f(
h^\ast )= {1\over 2}$.
Thereby, classical undecidability is recovered.
Stated pointedly: With regards to the question of
whether or not a computer halts,
the ``solution'' $h^\ast$
is equivalent to the throwing of a fair coin.

Therefore, the advance of quantum recursion theory over classical
recursion theory is not so much classical problem solving but {\em the
consistent representation of statements} which would give rise to
classical paradoxes.

\subsubsection{Proper quantum diagonalization}
The above argument used the continuity of qbit states as compared to the
two cbit states for a construction of fixed points of the
diagonalization operator. One could proceed a step further and allow
{\em nonclassical diagonalization procedures}. Such a step, albeit
operationalizable, has no classical operational equivalent, and thus no
classical interpretation.

Consider the entire range of twodimensional unitary transformations
\cite{murnaghan}
\begin{equation}
U(2)(\omega ,\alpha ,\beta ,\varphi )=e^{-i\,\beta}\,
\left(
\begin{array}{cc}
{e^{i\,\alpha }}\,\cos \omega
&
{-e^{-i\,\varphi }}\,\sin \omega
\\
{e^{i\,\varphi }}\,\sin \omega
&
{e^{-i\,\alpha }}\,\cos \omega
 \end{array}
\right)
 \quad ,
\end{equation}
where $-\pi \le \beta ,\omega \le \pi$,
$-\, {\pi \over 2} \le  \alpha ,\varphi \le {\pi \over 2}$, to act on
the qbit.
A typical example of a nonclassical operation on a qbit is
the ``square root of not'' gate
($
\sqrt{{\tt not}}
\sqrt{{\tt not}} =D$)
\begin{equation}
\sqrt{{\tt not}} =
{1 \over 2}
\left(
\begin{array}{cc}
1+i&1-i
\\
1-i&1+i
 \end{array}
\right)
\quad .
\end{equation}
Not all these unitary transformations have eigenvectors
associated with eigenvalues $1$ and thus fixed points.
Indeed, it is not difficult to see that only
unitary transformations of the form
\begin{equation}
[U(2)(\omega ,\alpha ,\beta ,\varphi )]^{-1}\,\mbox{diag}(1, e^{i\lambda
}) U(2)(\omega ,\alpha ,\beta ,\varphi )=
\left(
\begin{array}{cc}
{{\cos \omega }^2} + {e^{i\,\lambda }}\,{{\sin \omega }^2}&
{{{
{-1 + {e^{i\,\lambda
}}\over 2}
e^{-i\,\left(\alpha +\varphi \right) }}\,
\, \sin (2\,\omega )}} \\
{ -1 + {e^{i\,\lambda }}\over 2}
 {{{e^{i\,\left(\alpha
+\varphi \right) }}\,
 \sin
(2\,\omega )}}&
{e^{i\,\lambda }}\,{{\cos \omega }^2} + {{\sin
\omega }^2}
 \end{array}
\right)
\end{equation}
have fixed points.

Applying nonclassical operations on qbits with no fixed points
\begin{eqnarray}
D' &=&
[U(2)(\omega ,\alpha ,\beta ,\varphi )]^{-1}\,\mbox{diag}( e^{i\mu } ,
e^{i\lambda }) U(2)(\omega ,\alpha ,\beta ,\varphi )\nonumber \\ &=&
\left(
\begin{array}{cc}
  {e^{i\,\mu }}\,{{\cos (\omega )}^2} +
     {e^{i\,\lambda }}\,{{\sin (\omega )}^2}&
    {{{e^{-i\,\left( \alpha  + p \right) }\over 2}}\,
         \left( {e^{i\,\lambda }} - {e^{i\,\mu }} \right) \,\sin
(2\,\omega )}
       \\
{{{e^{i\,\left( \alpha  + p \right) }\over 2}}\,
        \left( {e^{i\,\lambda }} - {e^{i\,\mu }}  \right) \,\sin
(2\,\omega )}
       &{e^{i\,\lambda }}\,{{\cos (\omega )}^2} +
     {e^{i\,\mu }}\,{{\sin (\omega )}^2}
 \end{array}
\right)
\end{eqnarray}
with $\mu ,\lambda \neq n\pi$, $n\in {\Bbb N}_0$ gives rise to
eigenvectors which are not fixed points, but which acquire nonvanishing
phases $\mu , \lambda$ in the generalized diagonalization process.

\section{Quantum algorithmic information}

Quantum algorithmic information theory can be developed in analogy to
algorithmic information theory \cite{chaitin,calude,vitany}.
Before proceeding, though, one decisive strategic decision concerning
the physical character of
the program has to be made. This amounts to a restriction to purely {\em
classical} prefix-free programs.

The reason for classical programs, as well as for the requirement of
instant decodability, is the desired convergence of the Kraft sum
over the exponentially weighted program length
$\sum_p \exp ({\vert p\vert \log k})\le 1$, where
$\vert p\vert$ stands for the length of $p$ and $k$ is the base of the
code (for binary code, $k=2$).
If arbitrary qbits were allowed as program code, then the Kraft sum
would diverge.

Nevertheless, qbits are allowed as output. Since they are objects
defined in Hilbert space ${\cal H}$, the basic definitions of
algorithmic information theory have to be slightly adapted.

 The {\em canonical program} associated with an object $s\in {\cal H}$
 representable as vector in a Hilbert space ${\cal H}$ is denoted by
$s^\ast$ and defined by
 \begin{equation} s^\ast =\min_{C(p)=s}p\quad .
 \end{equation}
 I.e., $s^\ast$ is the first element in the ordered
 set of all strings that is a program for $C$ to calculate $s$.
 The string $s^\ast$ is thus the code of the smallest-size program
 which, implemented on a quantum computer, outputs $s$.
 (If several binary programs of equal length exist,
 the one is chosen which comes first in an enumeration using the usual
 lexicographic order relation ``$0<1$.'')

 Let again ``$\vert x \vert$'' of an object encoded as (binary) string
 stand for the length of that string.
The {\em quantum algorithmic information} $H(s)$ of an object $s\in
{\cal H}$
 representable as  vector in a Hilbert space ${\cal H}$ is defined as
the length
of the shortest  program $p$ which runs on a quantum computer
 $C$ and generates
the output $s$:
 \begin{equation}
H(s)=\vert s^\ast \vert =\min_{C(p)=s} \vert p\vert \quad .
 \end{equation}
If no program makes computer $C$ output $s$, then $H(s)=\infty $.

 The {\em joint quantum algorithmic information}
 $H(s,t)$ of two objects   $s\in {\cal H}$ and $t\in {\cal H}$
representable as vectors in a Hilbert space ${\cal H}$
 is the length of the smallest-size binary program to calculate
 $s$ and $t$ simultaneously.

 The {\em relative or conditional quantum algorithmic information}
 $H(s\vert t)$ of  $s\in {\cal H}$ given $t\in {\Bbb N}$
 is the length of the
 smallest-size binary program to calculate $s$ from a
{\em  smallest-size program} for $t$:
 \begin{equation}
H(s\vert t)=\min_{C(p,t^\ast )=s} \vert p\vert \quad .
 \end{equation}

Most features and results of algorithmic information theory hold for
quantum algorithmic information as well. In particular, we restrict our
attention to universal quantum computers whose quantum algorithmic
information content is machine-independent, such that the quantum
algorithmic information content of an arbitrary object does not exceed a
constant independent of that object. That is, for all objects $s\in
{\cal H}$ and two computers $C$ and $C'$ of this class,
 \begin{equation}
\vert H_C-H_{C'}\vert
=O(1)\quad . \label{machine-independece}
 \end{equation}

Furthermore, let $s$ and $t$ be two objects representable as vectors in
Hilbert space. Then (recall that $t\in {\Bbb N}$),
 \begin{eqnarray}
 H(s,t)&=& H(t,s)+O(1)\quad ;\\
 H(s\vert s)&=&O(1)\quad ;\\
 H(H(s)\vert s)&=&O(1)\quad ;\\
 H(s)&\le & H(s,t)+O(1)\quad ;\\
 H(s\vert t)&\le &H(s)+O(1)\quad ;\\
 H(s,t)&= & H(s)+H(t\vert s^\ast )+O(1) \quad \mbox{(if }s^\ast \mbox{
is classical)}
\label{subadditivity1} \quad ;\\
 H(s,t)&\le & H(s)+H(t)+O(1) \quad {\rm
 (subadditivity)}\label{subadditivity} \quad ;\\
 H(s,s)&=&H(s)+O(1)\quad ;\\
 H(s,H(s))&=&H(s)+O(1)\quad .
 \end{eqnarray}

Notice
that there exist sets of objects
$S=\{s_1,\ldots
,s_n\}$, $n<\infty$
whose algorithmic information content $H(S)$ is arbitrary small compared
to the algorithmic information content of some unpecified {\em single}
elements $s_i\in S$; i.e.,
\begin{equation}
H(S)<\max_{s_i\in S}H(s_i)\quad .
\end{equation}

\section{Quantum omega}
Chaitin's $\Omega$ \cite{chaitin,solovay,calude} is a magic
number.
It is a measure for arbitrary programs to take a finite number of
execution steps and
then halt.
It contains the solution of all halting problems, and hence of
questions codable into halting problems, such as Fermat's theorem.
It contains the solution of the question of whether or not a particular
exponential Diophantine equation
has infinitely many or a finite number of solutions.
And, since $\Omega$ is provable ``algorithmically incompressible,'' it
is Martin-L\"of/Chaitin/Solovay random. Therefore, $\Omega$ is both: a
mathematicians ``fair coin,''  and a formalist's nightmare.

Here, $\Omega$ is generalized to quantum computations.\footnote{The
quantum omega was invented in a meeting of G. Chaitin,
A. Zeilinger and the author in a Viennese coffee house
(Caf\'{e} Br\"aunerhof) in January 1991.
Thus, the group should be credited for the original invention,
whereas any blame should remain with the author.}

In the orthonormal halting basis $\{ t  ,f
\}$,
the computer $C$ with
classical input
$p_i$ can be represented by
$
C(\tau ,p_i)    = t
\,(
t,
 C(\tau ,p_i))
+
f\,(f,C(\tau ,p_i))$.

Recall
that initially,
i.e., at time $\tau =0$, the halting bit
is in a coherent 50:50-superposition; i.e.,
in terms of the halting basis,
$C(0,p_i)    =
( t+f)/ \sqrt{2}$
for all $p_i\in A^\ast$.
This corresponds to the fact that initially it is unknown whether or not
the computer halts on $p_i$.
When during the time evolution the  computer has completed its task,
the halting bit value is switched to $t$
by some internal operation.
If the computer never halts,
the halting bit value is switched to $f$
by some internal operation.
Otherwise it
remains in the  coherent 50:50-superposition.

Alternatively, the computer could be initially prepared
in the non-halting state $f $.
After completion of the
task, the halting bit is again
switched to
the halting state $ t$.

In analogy to the fully classical case \cite{chaitin,solomonoff,calude},
the {\em quantum halting amplitude}\footnote{
The definition of $\Omega$ and $\Upsilon$ differ slightly from the ones
introduced by the author previously \cite{quantum-omega}.
}
 $\Omega $ can be defined
 as a weighted expectation over all computations of $C$ with classical
input
$p_i$ ($\vert p_i\vert $ stands for the length of $p_i$)
 \begin{equation}
   \Omega \equiv
\sum_{C(p_i)\in {\cal H}}
2^{-\vert p_i\vert/2
}
 (  t,   C(p_i) )
\quad .
\label{e:qo}
 \end{equation}

Likewise, the halting amplitude for a particular output state
$ s
$,
 \begin{equation}
  \Upsilon ( s )  \equiv
\sum_{ C(p_i) = s }
2^{-\vert p_i\vert /2
}
( t,C(p_i) )
\quad .
\label{e:qop}
 \end{equation}
For a set of output states
 $S=\{
s_1 ,
s_2 ,
s_3 ,
\ldots
, s_n\}$
 which correspond to mutually orthogonal
vectors in Hilbert space,
 \begin{equation}
  \Upsilon ( S )  \equiv
\sum_{ C(p_i) \in S }
2^{-\vert p_i\vert   /2
}
( t,  C(p_i) )
\quad .
\label{e:qop1}
 \end{equation}

Terms corresponding to different programs and states have to be summed
up incoherently.
Thus, the corresponding probabilities are
 \begin{eqnarray}
\vert \Omega \vert^2&=& \sum_{C(p_i)\in {\cal  H}}
2^{-\vert p_i\vert
}
\vert (  t,   C(p_i) ) \vert^2\\
P(s)&\equiv &
\vert \Upsilon
(s)\vert^2 =
\sum_{C(p_i)=s}
2^{-\vert p_i\vert
}
\vert (  t,   C(p_i) ) \vert^2\\
P(S)&\equiv &
\sum_{C(p_i)\in S}
\vert \Upsilon
(s)\vert^2 =
\sum_{C(p_i)\in S}
2^{-\vert p_i\vert
}
\vert (  t,   C(p_i) ) \vert^2
\quad .
 \end{eqnarray}

The following relations hold,
 \begin{eqnarray}
  \Upsilon ( S )  &=&\sum_{s_i\in S} \Upsilon (s_i)\quad ,\\
\Omega  &=&\Upsilon( {\cal H})=\sum_{s_i\in {\cal  H}} \Upsilon
(s_i)\quad .
 \end{eqnarray}
For $s\subset S\subset {\cal H}$,
\begin{equation}
0\le P(s)\le P(S)\le \vert \Omega \vert^2 \le 1
\quad .
\label{e:qop11}
\end{equation}

Alternatively, the {\em quantum halting probability} and the {\em
quantum algorithmic information} by the quantum algorithmic information
content. That is,
\begin{eqnarray}
 P^\ast (s)&=&2^{-\vert s^\ast \vert }=2^{-H(s)} \label{p-ast}\\
 P^\ast (S)&=&\sum_{s_i\in S}P^\ast (s)=\sum_{s\in S}2^{-H(s)}\\
 P^\ast ({\cal H} )= \vert \Omega^\ast \vert^2 &=& \sum_{n\in {\cal H} }
 2^{-H(n)}\label{omega-ast} \quad  .
 \end{eqnarray}

 \begin{eqnarray}
\vert \Omega^\ast \vert^2
&\le &   \vert \Omega \vert^2
\quad ,\\
P^\ast (s)
&\le &
P (s)
\quad ,\\
P^\ast (S)
&\le &
P (S)
\quad .
 \end{eqnarray}

The following relations are either a direct consequence of the
definition (\ref{p-ast}) or follow from the fact that for programs in
prefix code, the algorithmic probability is concentrated on the minimal
size programs, or alternatively, that there are few minimal programs:
 \begin{eqnarray}
 H(s)&=&-\log_2 P^\ast (s)\label{h-p2}\quad ;\\
 H(s)&=&-\log_2 P(s)+O(1)\label{h-p}\quad .
 \end{eqnarray}

Notice again that, because of complementarity, single qbits cannot be
determined precisely. They just appear experimentally as some clicks in
a counter. What we can effectively do is to observe a successive
number of such qbits, one after the other, from ``similar'' computation
processes (same preparation, same evolution). By performing these
measurements on
``similar'' qbits, one can
``determine'' this qbit within an
$\varepsilon$-neighborhood
only.

For nontrivial choices of the quantum computer $C$,
several remarks are in order.
(In what follows, we mention only $\Omega$, but the comments apply to
$\Upsilon$ as well.)
If the program is also
coded in qbits, the above sum becomes an
integral over continuously many states per code symbol of the programs.
In this case, the Kraft sum needs not converge.
Just as for the classical analogue it is possible to ``compute''
$\Omega $ as a limit from below
by considering
in the $t$'th computing step (time $\tau$) all programs of length
$\tau$ which have already halted.
(This ``computation'' suffers
from a radius of convergence which decreases slower than any recursive
function.)
The quantum $\Omega$ is complex.
$\vert \Omega \vert^2$  can be interpreted as a
measure for the halting probability of
$C$; i.e., the probability that an arbitrary (prefix-free) program
halts on
$C$.

Finally,
any irreversible measurement of $\vert \Omega \vert^2$
causes a state collapse.
Since $ C(\tau ,p_i) $ may not be in a
pure state, the  series in
 (\ref{e:qo}) and
 (\ref{e:qop})
will not be uniquely defined even for {\em finite} times.
Thus the {\em nondeterministic} character of $\Omega$ is not only based
on classical recursion theoretic arguments \cite{chaitin} but
also on the metaphysical assumption that  God plays the
quantum dice.

\appendix
\section*{Appendices}
 \addcontentsline{toc}{section}{Appendix}
\section{Two-state system}
\label{tws}

Having set the stage of the quantum formalism, an
elementary twodimensional example of a two-state system shall be
exhibited
(\cite{feynman-III}, p. 8-11). Let us denote the two base states by
$ 1 $ and $
2$.
Any arbitrary physical state $ \psi $ is a coherent
superposition of
$ 1 $ and $
2$
and can be written as
$ \psi  =
 1 ( 1, \psi ) +
2 ( 2, \psi )$ with the two coefficients
$( 1, \psi ) ,
( 2, \psi ) \in {\Bbb C}$.

Let us discuss two particular types of evolutions.

First, let us discuss the Schr\"odinger equation with
diagonal Hamilton matrix, i.e., with vanishing off-diagonal elements,
\begin{equation}
 H_{ij} =\left( \begin{array}{cc} E_1  & 0\\
0 &E_2
 \end{array} \right)
 \quad .
\end{equation}
In this case, the Schr\"odinger equation decouples and
reduces to
\begin{equation}
i\hbar {\partial \over \partial t} ( 1 , \psi (t) )   =
E_1 ( 1 , \psi (t) )
\quad ,\qquad
i\hbar {\partial \over \partial t} ( 2 , \psi  (t) )   =
E_2 ( 2 , \psi (t) )
\quad ,
\end{equation}
resulting in
\begin{equation}
( 1 , \psi (t) )   = a e^{-iE_1t/\hbar}
\quad ,\qquad
( 2 , \psi (t) )   = b e^{-iE_2t/\hbar}
\quad ,
\end{equation}
with $a,b\in {\Bbb C}$, $\vert a\vert ^2+\vert b\vert ^2=1$.
These solutions
correspond to {\em stationary states} which do not change in time; i.e.,
the probability to find the system in the two states is constant
\begin{equation}
\vert ( 1 , \psi  )  \vert^2 = \vert a\vert ^2
\quad ,\qquad
\vert ( 2 , \psi  )   \vert^2 = \vert b\vert^2
\quad .
\end{equation}

Second, let us discuss the Schr\"odinger equation with
with non-vanishing but equal off-diagonal elements $-A$
and with equal diagonal elements $E$
of the
Hamiltonian matrix; i.e.,
\begin{equation}
H_{ij} =\left( \begin{array}{cc} E  & -A\\
-A &E
 \end{array} \right)
 \quad .
\end{equation}
In this case, the Schr\"odinger equation
reads
\begin{eqnarray}
i\hbar {\partial \over \partial t} ( 1 , \psi (t)  )   &=&
E ( 1 , \psi (t) ) - A  ( 2 , \psi (t) )
\quad ,\\
i\hbar {\partial \over \partial t} ( 2 , \psi (t) )   &=&
E ( 2 , \psi (t) ) - A  ( 1 , \psi (t) )
\quad .
\end{eqnarray}
These equations can be solved in a number of ways. For example, taking
the sum and the difference of the two, one obtains
\begin{eqnarray}
i\hbar {\partial \over \partial t} (( 1 , \psi (t) ) +( 2
, \psi (t) ) )
&=&
(E-A) (( 1 , \psi (t) ) +  ( 2 , \psi (t) ))
\quad ,\\
i\hbar {\partial \over \partial t} (( 1 , \psi (t)  ) -( 2
, \psi  (t)) )
&=&
(E+A) (( 1 , \psi (t) ) -  ( 2 , \psi (t) ))
\quad .
\end{eqnarray}
The solution are again two stationary states
\begin{eqnarray}
( 1 , \psi (t) ) +( 2 , \psi (t) )
&=&
ae^{-(i/\hbar )(E-A)t}
\quad ,\\
( 1 , \psi  (t)) -( 2 , \psi  (t))
&=&
be^{-(i/\hbar )(E+A)t}
\quad .
\end{eqnarray}
Thus,
\begin{eqnarray}
( 1 , \psi  (t))
&=&
{a\over 2}e^{-(i/\hbar )(E-A)t}+
{b\over 2}e^{-(i/\hbar )(E+A)t}
\label{el:nh21}
\quad ,\\
( 2 , \psi  (t))
&=&
{a\over 2}e^{-(i/\hbar )(E-A)t}-
{b\over 2}e^{-(i/\hbar )(E+A)t}
\label{el:nh22}
\quad .
\end{eqnarray}

Assume now that initially, i.e., at $t=0$, the system was in state
$, 1) =, \psi (t=0))$. This assumption corresponds to
$
( 1 , \psi (t=0)  ) =1
$
and
$
( 2 , \psi (t=0)  ) =0
$.
What is the probability that the system will be found
in the state
$
2
$
at the time $t>0$, or that
it will still be found
in the state
$
 1
$
at the time $t>0$?
Setting $t=0$ in equations
(\ref{el:nh21}) and (\ref{el:nh22}) yields
\begin{equation}
( 1 , \psi  (t=0))   = {a+b\over 2}=1
\quad ,\qquad
( 2 , \psi  (t=0))   = {a-b\over 2}=0
\quad ,
\end{equation}
and thus $a=b=1$. Equations
(\ref{el:nh21}) and (\ref{el:nh22}) can now be evaluated at $t>0$ by
substituting
1 for
$a$ and
$b$,
\begin{eqnarray}
( 1 , \psi  (t))
&=&
e^{-(i/\hbar )Et}\left[
{
e^{(i/\hbar )At}+
e^{-(i/\hbar )At}
\over 2}
\right]
=
e^{-(i/\hbar )Et}\cos {At\over \hbar }
\label{el:nh23}
\quad ,\\
( 2 , \psi  (t))
&=&
e^{-(i/\hbar )Et}\left[
{
e^{(i/\hbar )At}-
e^{-(i/\hbar )At}
\over 2}
\right]
=
i\,e^{-(i/\hbar )Et}\sin {At\over \hbar }
\label{el:nh24}
\quad .
\end{eqnarray}
Finally, the probability that the system is in state
$, 1)$ and
$, 2)$
is
\begin{equation}
\vert ( 1 , \psi  (t)) \vert^2  = \cos^2{At\over \hbar}
\quad ,\qquad
\vert ( 2 , \psi  (t)) \vert^2  = \sin^2{At\over \hbar}
\quad ,
\label{eqn:p}
\end{equation}
respectively. This results in an oscillation of the transition
probabilities.

\begin{figure}
\begin{center}
\unitlength=0.5mm
\special{em:linewidth 0.4pt}
\linethickness{0.4pt}
\begin{picture}(178.25,103.33)
\put(156.25,19.67){\circle{14.00}}
\put(156.25,19.67){\makebox(0,0)[cc]{H}}
\put(101.25,39.67){\circle{14.00}}
\put(101.25,39.67){\makebox(0,0)[cc]{H}}
\put(171.25,59.67){\circle{14.00}}
\put(171.25,59.67){\makebox(0,0)[cc]{H}}
\put(129.25,89.67){\circle{14.00}}
\put(129.25,89.67){\makebox(0,0)[cc]{N}}
\put(107.25,35.67){\line(3,-1){42.00}}
\put(160.25,25.33){\line(1,3){9.33}}
\put(106.92,43.67){\line(4,1){57.33}}
\put(103.25,46.33){\line(3,5){22.33}}
\put(131.25,82.67){\line(2,-5){22.67}}
\put(135.58,87.33){\line(5,-4){29.67}}
\put(65.00,49.67){\circle{14.00}}
\put(65.00,49.67){\makebox(0,0)[cc]{H}}
\put(10.00,69.67){\circle{14.00}}
\put(10.00,69.67){\makebox(0,0)[cc]{H}}
\put(80.00,89.67){\circle{14.00}}
\put(80.00,89.67){\makebox(0,0)[cc]{H}}
\put(16.00,65.67){\line(3,-1){42.00}}
\put(69.00,55.33){\line(1,3){9.33}}
\put(15.67,73.67){\line(4,1){57.33}}
\put(37.58,19.67){\circle{14.00}}
\put(37.58,19.67){\makebox(0,0)[cc]{N}}
\put(37.50,5.00){\line(0,1){7.50}}
\put(37.50,26.67){\line(0,1){60.83}}
\put(128.75,18.33){\line(0,1){64.17}}
\put(128.75,96.67){\line(0,1){6.67}}
\put(12.92,62.92){\line(1,-2){19.17}}
\put(43.33,23.75){\line(1,1){18.75}}
\put(75.00,84.17){\line(-3,-5){34.58}}
\put(163.39,88.81){\makebox(0,0)[cc]{$\mid 2\rangle$}}
\put(14.17,89.58){\makebox(0,0)[cc]{$\mid 1\rangle$}}
\end{picture}
\end{center}
\caption{The two equivalent geometric arrangements of the ammonia
(NH$_3$) molecule.
 \label{f:ammonia}}
\end{figure}
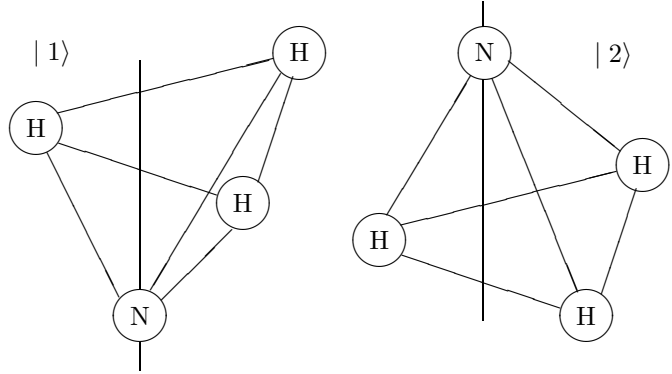

Let us shortly mention one particular realization of a two-state system
which, among many others, has been discussed in the Feynman lectures
 \cite{feynman-III}.
Consider an ammonia (NH$_3$) molecule. If one fixes the plane spanned by
the three hydrogen atoms, one observes two possible spatial
configurations $, 1)$ and $, 2)$, corresponding to position
of the nitrogen atom in the lower or the upper hemisphere, respectively
(cf. Fig.
 \ref{f:ammonia}).
The nondiagonal elements of the Hamiltonian $H_{12}=H_{21}=-A$
correspond to a nonvanishing transition probability from one such
configuration into the other.
If the ammonia has been originally in state $, 1)$, it will
constantly swing back and forth between the two states, with a
probability given by equations
(\ref{eqn:p}).

\section{From single to multiple quanta --- ``second'' field
quantization}
\label{a:2}

The quantum formalism introduced in the main text is about {\em single}
quantized
objects. What if one wants to consider many such objects? Do we have to
add assumptions in order to treat such multi-particle, multi-quanta
systems appropriately?

The answer is yes. Experiment and theoretical reasoning (the
representation theory of the Lorentz group \cite{urbantke} and the
spin-statistics
 theorem \cite{jauch-rohrlich,lurie,bogol-shi,itzy}) indicate that
there are
(at least) two basic types of states (quanta, particles):
{\em bosonic} and
{\em fermionic} states.
Bosonic states have what is called ``integer spin;'' i.e.,
$s_b=0,\hbar ,2 \hbar ,3\hbar ,\ldots$, whereas fermionic states have
``half-integer spin;''
$s_f={1\hbar \over 2},{3\hbar \over2},{5\hbar \over 2}\ldots$.
Most important, they are characterized by the way identical copies of
them can be ``brought together.''
Consider two boxes, one for identical bosons, say photons, the other one
for identical fermions, say electrons.
For the first, bosonic, box, the probability that another identical
boson is added {\em increases with the number of identical bosons} which
are already in the box. There is a tendency of bosons to ``condensate''
into the same state.
The second, fermionic box, behaves quite differently. If it is already
occupied by one fermion, another identical fermion cannot enter. This is
expressed in the {\em Pauli exclusion principle:}
A system of fermions can never occupy a configuration of individual
states in which two individual states are identical.

How can the bose condensation and the Pauli exclusion principle be
implemented? There are several forms of implementation (e.g.,
fermionic behavior via Slater-determinants), but the most compact and
widely practiced form uses operator algebra. In the following we shall
present this formalism in the context of quantum field theory
\cite{har,lipkin,jauch-rohrlich,lurie,bogol-shi,itzy,glauber}.

A {\em classical} field can be represented by its Fourier transform
(``$\ast$'' stands for complex conjugation)
\begin{eqnarray}
A(x,t)&=&A^{(+)}(x,t)
+
A^{(-)}(x,t)\\
A^{(+)}(x,t)
&=&
[A^{(-)}(x,t)]^\ast\\
A^{(+)}(x,t)
&=&
\sum_{k_i,s_i} a_{k_i,s_i}u_{k_i,s_i}(x)e^{-i\omega_{k_i} t}\quad ,
\label{el:a1}
\end{eqnarray}
where $\nu =\omega_{k_i}/2\pi$ stands for the frequency in the field
mode
labeled by  momentum $k_i$ and
$s_i$ is some observable such as spin or
polarization.
$u_{k_i,s_i}$ stands for the polarization vector
(spinor) at $k_i,s_i$,
and, most important with regards to the quantized case, {\em
complex-valued} Fourier coefficients $a_{k_i,s_i}\in {\Bbb C}$.

{}From now on, the $k_i,s_i$-mode will be abbreviated by the symbol $i$;
i.e.,
$1 \equiv k_1,s_1$,
$2 \equiv k_2,s_2$,
$3 \equiv k_3,s_3$, $\ldots$,
$i \equiv k_i,s_i$, $\ldots$.

In (second\footnote{of course, there is only ``the one and only''
quantization,
the term ``second'' often refers to operator techniques for multiquanta
systems; i.e.,  quantum field theory}) quantization,
the classical Fourier coefficients $a_{i}$ become re-interpreted as {\em
operators}, which obey the following algebraic rules
(scalars would not do the trick).
For {\em bosonic} fields (e.g., for the electromagnetic field),
the {\em commutator} relations are
(``$\dagger$'' stands for self-adjointness):
\begin{eqnarray}
\left[a_{i},a_{j}^\dagger \right]&=&
a_{i} a_{j}^\dagger - a_{j}^\dagger a_{i}
=\delta_{ij} \quad , \\
\left[
a_{i}
,
a_{j}
\right]
&=&
\left[
a_{i}^\dagger
,
a_{j}^\dagger
\right]
= 0 \quad .
\end{eqnarray}

For {\em fermionic} fields (e.g., for the electron field),
the {\em anti-commutator} relations are:
\begin{eqnarray}
\{
a_{i}
,
a_{j}^\dagger
\}
&=&
a_{i}
a_{j}^\dagger
+
a_{j}^\dagger
a_{i}
=\delta_{ij} \quad ,  \\
\{
a_{i}
,
a_{j}
\}
&=&
\{
a_{i}^\dagger
,
a_{j}^\dagger
\}
= 0 \quad .
\end{eqnarray}
The anti-commutator relations, in particular
$
\{
a_{j}^\dagger
,
a_{j}^\dagger
\} =
2(
a_{j}^\dagger)^2=0
$,
are just a formal
expression of the Pauli exclusion
principle stating that, unlike bosons, two or more identical
fermions cannot co-exist.

The operators
$
a_{i}^\dagger
$
and
$
a_{i}
$
are called {\em creation} and {\em annihilation} operators,
respectively. This terminology suggests itself if one introduces
{\em Fock states} and
the {\em occupation number formalism}.
$
a_{i}^\dagger
$
and
$
a_{i}
$
are applied to Fock states to following effect.

The Fock space associated with a quantized field will be
the direct product
of all Hilbert spaces ${\cal H}_i$; i.e.,
\begin{equation}
\prod_{i\in {\Bbb I}} {\cal H}_i
\quad ,
\end{equation}
where ${\Bbb I}$ is an index set characterizing all different
field modes labeled by $i$.
Each boson (photon) field mode is equivalent to a harmonic oscillator
\cite{glauber,loudon};
each fermion (electron, proton, neutron) field mode is equivalent to the
Larmor precession of an electron spin.

In what follows, only finite-size systems
are studied.
The Fock states are based upon the Fock vacuum.
The Fock vacuum is a direct product of states $\mid  0_i\rangle $
of the $i$'th Hilbert space ${\cal H}_i$ characterizing mode $i$; i.e.,
\begin{eqnarray}
\mid 0\rangle
&=& \prod_{i\in {\Bbb I}} \mid  0 \rangle_i
=
\mid  0 \rangle_1 \otimes
\mid  0 \rangle_2 \otimes
\mid  0 \rangle_3 \otimes
\cdots
\nonumber \\
&=&\mid \bigcup_{i\in {\Bbb I}} \{0_{i}\} \rangle
=
\mid \{
0_{1},
0_{2},
0_{3},
\ldots
\}\rangle
\quad ,
\end{eqnarray}
where again ${\Bbb I}$ is an index set characterizing all different
field modes labeled by $i$.
``$0_{i}$'' stands for $0$ (no) quantum (particle)
in the state characterized by the quantum numbers ${i}$.
Likewise, more generally,
``$N_{i}$'' stands for $N$  quanta (particles)
in the state characterized by the quantum numbers ${i}$.

The annihilation operators
$
a_{i}
$
are designed to destroy one quantum (particle) in state ${i}$:
\begin{eqnarray}
&&a_{j} \mid 0\rangle
=0\quad ,\\
&&a_{j}
\mid \{
0_{1},
0_{2},
0_{3},
\ldots ,
0_{{j-1}},
N_{j},
0_{{j+1}}, \ldots
\}\rangle
= \nonumber \\
&&\qquad = \sqrt{N_{j}}
\mid \{
0_{1},
0_{2},
0_{3},
\ldots ,
0_{{j-1}},
(N_{j}-1),
0_{{j+1}}, \ldots
\}\rangle
\quad .
\end{eqnarray}

The creation operators
$
a_{i}^\dagger
$
are designed to create one quantum (particle) in state ${i}$:
\begin{equation}
a_{j}^\dagger
\mid 0 \rangle
=
\mid \{
0_{1},
0_{2},
0_{3},
\ldots ,
0_{{j-1}},
1_{j},
0_{{j+1}}, \ldots
\}\rangle
\quad .
\end{equation}
More generally,
$N_{j}$ operators
$(a_{j}^\dagger )^{N_{j}}$
 create an $N_{j}$-quanta
(particles) state
\begin{equation}
(a_{j}^\dagger )^{N_{j}}
\mid 0\rangle
\propto
\mid \{
0_{1},
0_{2},
0_{3},
\ldots ,
0_{{j-1}},
N_{j},
0_{{j+1}}, \ldots
\}\rangle
\quad .
\end{equation}
For fermions, $N_{j}\in \{ 0,1\}$ because of the Pauli exclusion
principle. For bosons, $N_{j}\in {\Bbb N}_0$.
With proper normalization [which can motivated by
the
(anti-)commutator relations and by  $\vert ( X, X)
\vert^2
=1$], a state
$
$
containing
$
N_{1}
$
quanta (particles) in mode ${1}$,
$
N_{2}
$
quanta (particles) in mode ${2}$,
$
N_{3}
$
quanta (particles) in mode ${3}$, {\it etc.}, can be generated from
the Fock vacuum by
\begin{equation}
\mid \bigcup_{ i\in {\Bbb I}} \{N_{i}\} \rangle
\equiv
\mid \{
N_{1},
N_{2},
N_{3},
\ldots
\}\rangle
=
\prod_{i\in {\Bbb I}}{(a_{i}^\dagger )^{N_{i}}\over
\sqrt{N_{i}!}}
\mid 0 \rangle
\quad .
\label{el:a3}
\end{equation}

As
has been stated by Glauber (\cite{glauber}, p. 64),
\begin{quote}
$\ldots$ in quantum theory, there is an  infinite set of complex
numbers which
specifies the state of a single mode. This is in contrast to classical
theory where each mode may be described by a  single complex
number. This shows that there is vastly more freedom in quantum theory
to invent states of the world than there is in the classical
theory. We cannot think
of quantum theory and classical theory in one-to-one terms at all. In
quantum theory, there exist whole spaces which have no classical
analogues, whatever.
\end{quote}

\section{Quantum interference}
\label{a:toolbox}

In what follows,
we shall make use of a simple ``toolbox''-scheme of combining
lossless elements of an experimental setup for the theoretical
calculation
\cite{green-horn-zei}.
 The
elements of this ``toolbox'' are listed in Table \ref{ta:1}.
These ``toolbox'' rules can be rigorously motivated by the full quantum
optical calculations (e.g.,  \cite{yurke-86,teich:90})
but are much easier to use.
In what follows, the factor $i$ resulting from a phase shift of $\pi /2$
associated with the reflection at a mirror $M$ is {\em omitted.}
However, at a half-silvered mirror beam splitter, the relative factor
$i$ resulting from a phase shift of $\pi /2$ {\em is} kept.
(A detailed calculation \cite{born-wolf} shows that
this phase shift of $\pi /2$ is an approximation which is exactly valid
only for particular system parameters).
$T$ and $R=\sqrt{1-T^2}$ are transmission and reflection coefficients.
Notice that the ``generic''
beam splitter can  be realized by a
half-silvered mirror and a successive phase shift of $\varphi =-\pi/2
$ in the reflected channel; i.e.,
$ a  \rightarrow
( b  +i c
)/\sqrt{2}
 \rightarrow
( b  +ie^{-i\pi /2} c
)/\sqrt{2}
 \rightarrow
( b  + c
)/\sqrt{2}
$.
Note also that, in the ``second quantization'' notation, for $i<j$,
\begin{equation}
\mid  i\rangle  \mid  j\rangle  \equiv
a_i^\dagger a_j^\dagger \mid  0\rangle  =
\mid  i\rangle  \otimes \mid  j\rangle  =
\mid  0_1,0_2,0_3,\ldots
,0_{i-1},1_i,0_{i+1},
\ldots, 0_{j-1},1_j,0_{j+1},\ldots \rangle
\quad .
\end{equation}
In present-day quantum optical nonlinear devices (NL), parametric
up- or down-conversion, i.e., the production of a single quant
(particle) from two field quanta (particles) and the production of two
field quanta (particles) from a single one occurs at the very
low amplitude rate of $\eta \approx 10^{-6}$.
\begin{table}
\begin{tabular}{l|c|lllllll}
\hline
\hline
physical process & symbol & state transformation
\\
\hline
\hline
reflection at mirror
&&
$ a  \rightarrow  b  =i a  $
\\
&
\unitlength 0.70mm
\linethickness{0.4pt}
\begin{picture}(26.00,26.00)
\put(4.00,21.67){\line(1,0){18.00}}
\multiput(18.00,25.67)(0.12,-0.12){67}{\line(0,-1){0.12}}
\put(21.67,21.67){\line(0,-1){16.67}}
\put(11.00,25.67){\makebox(0,0)[cc]{$a$}}
\put(25.67,10.00){\makebox(0,0)[cc]{$b$}}
\put(24.67,26.00){\makebox(0,0)[cc]{$M$}}
\end{picture}
\\
\hline

``generic'' beam splitter
&&
$ a  \rightarrow ( b  + c
)/\sqrt{2}$
\\
&
\unitlength=0.70mm
\special{em:linewidth 0.4pt}
\linethickness{0.4pt}
\begin{picture}(40.00,17.00)
\put(5.00,10.00){\line(1,0){10.00}}
\put(15.00,5.00){\dashbox{2.00}(10.00,10.00)[cc]{}}
\put(25.00,10.00){\line(3,-1){15.00}}
\put(25.00,10.00){\line(3,1){15.00}}
\put(9.00,14.00){\makebox(0,0)[cc]{$a$}}
\put(34.00,17.00){\makebox(0,0)[cc]{$b$}}
\put(34.00,2.00){\makebox(0,0)[cc]{$c$}}
\end{picture}
&
\\
\hline

transmission/reflection
&&
$ a  \rightarrow ( b  +i c
)/\sqrt{2}$
\\
by a beam splitter &&
$ a  \rightarrow T b  +iR c
$,
\\
(half-silvered mirror)
&
&$T^2+R^2=1$, $T,R\in [0,1]$
\\
&
\unitlength 0.70mm
\linethickness{0.4pt}
\begin{picture}(44.00,34.34)
\put(4.00,28.34){\line(1,0){40.00}}
\put(14.00,28.34){\line(0,-1){25.00}}
\put(28.00,31.34){\makebox(0,0)[cc]{$b$}}
\put(17.00,15.34){\makebox(0,0)[cc]{$c$}}
\put(14.00,34.34){\makebox(0,0)[cc]{$S_1$}}
\multiput(13.00,29.34)(0.12,-0.12){17}{\line(0,-1){0.12}}
\multiput(12.00,30.34)(-0.12,0.12){17}{\line(0,1){0.12}}
\multiput(16.00,26.34)(0.12,-0.12){17}{\line(0,-1){0.12}}
\put(7.00,24.34){\makebox(0,0)[cc]{$a$}}
\end{picture}
\\
\hline

phase-shift $\varphi$&
&
$ a  \rightarrow  b =  a e^{i
\varphi
}$
\\
&
\unitlength=0.70mm
\special{em:linewidth 0.4pt}
\linethickness{0.4pt}
\begin{picture}(26.00,14.34)
\put(14.00,14.34){\makebox(0,0)[cc]{$\varphi$}}
\put(3.67,11.34){\makebox(0,0)[cc]{$a$}}
\put(24.33,12.00){\makebox(0,0)[cc]{$b$}}
\put(1.00,7.00){\line(1,0){25.00}}
\put(9.00,4.00){\framebox(9.00,6.00)[cc]{}}
\end{picture}
\\
\hline

parametric down-conversion&
&
$ a  \rightarrow  \eta  b\rangle   c $
\\
&
\unitlength 0.70mm
\linethickness{0.4pt}
\begin{picture}(45.00,22.00)
\put(17.00,10.00){\framebox(10.00,10.00)[cc]{$NL$}}
\multiput(27.00,15.00)(0.31,0.12){59}{\line(1,0){0.31}}
\multiput(45.00,8.00)(-0.31,0.12){59}{\line(-1,0){0.31}}
\put(4.00,15.00){\line(1,0){13.00}}
\put(31.00,21.00){\makebox(0,0)[cc]{$b$}}
\put(31.00,9.00){\makebox(0,0)[cc]{$c$}}
\put(10.00,18.67){\makebox(0,0)[cc]{$a$}}
\end{picture}
\\
\hline

parametric up-conversion&
&
$ a\rangle  \mid  b  \rightarrow  \eta  c $
\\
&
\unitlength 0.70mm
\linethickness{0.4pt}
\begin{picture}(41.34,17.33)
\put(20.67,5.33){\framebox(10.00,10.00)[cc]{$NL$}}
\multiput(3.67,17.33)(0.29,-0.12){59}{\line(1,0){0.29}}
\multiput(20.67,10.33)(-0.29,-0.12){59}{\line(-1,0){0.29}}
\put(36.00,14.00){\makebox(0,0)[cc]{$c$}}
\put(14.67,15.33){\makebox(0,0)[cc]{$a$}}
\put(14.67,4.33){\makebox(0,0)[cc]{$b$}}
\put(30.67,10.33){\line(1,0){10.67}}
\end{picture}
\\
\hline

amplification
&&
$ A_i  a  \rightarrow  \vert b ;G,N\rangle   $
\\
&
\unitlength=0.70mm
\special{em:linewidth 0.4pt}
\linethickness{0.4pt}
\begin{picture}(26.00,14.34)
\put(14.00,14.34){\makebox(0,0)[cc]{$G,N$}}
\put(3.67,11.34){\makebox(0,0)[cc]{$a$}}
\put(24.33,12.00){\makebox(0,0)[cc]{$b$}}
\put(1.00,7.00){\line(1,0){25.00}}
\put(9.00,4.00){\framebox(9.00,6.00)[cc]{}}
\put(9.00,4.00){\rule{9.00\unitlength}{6.00\unitlength}}
\end{picture}
\\
\hline
\hline
\end{tabular}
\caption{``Toolbox'' of lossless elements for quantum interference
devices.
\label{ta:1}}
\end{table}

\begin{figure}
\begin{center}
\unitlength 0.70mm
\linethickness{0.4pt}
\begin{picture}(83.67,61.00)
\put(62.67,40.00){\line(0,-1){25.00}}
\put(10.00,55.00){\makebox(0,0)[cc]{$L$}}
\put(10.00,55.00){\circle{10.00}}
\put(15.00,55.00){\line(1,0){40.00}}
\put(44.67,55.00){\line(1,0){18.00}}
\put(25.00,55.00){\line(0,-1){25.00}}
\put(25.00,30.00){\line(1,0){13.00}}
\put(62.67,55.00){\line(0,-1){25.00}}
\put(62.67,30.00){\line(-1,0){35.00}}
\put(62.67,30.00){\line(1,0){13.00}}
\put(62.67,30.00){\line(0,-1){13.00}}
\multiput(58.67,59.00)(0.12,-0.12){67}{\line(0,-1){0.12}}
\multiput(21.00,35.00)(0.12,-0.15){67}{\line(0,-1){0.15}}
\put(80.17,30.00){\oval(7.00,8.00)[r]}
\put(83.67,36.00){\makebox(0,0)[cc]{$D_1$}}
\put(28.00,42.00){\makebox(0,0)[cc]{$c$}}
\put(25.00,61.00){\makebox(0,0)[cc]{$S_1$}}
\put(56.67,38.00){\makebox(0,0)[cc]{$S_2$}}
\multiput(24.00,56.00)(0.12,-0.12){17}{\line(0,-1){0.12}}
\multiput(23.00,57.00)(-0.12,0.12){17}{\line(0,1){0.12}}
\multiput(27.00,53.00)(0.12,-0.12){17}{\line(0,-1){0.12}}
\multiput(61.67,31.00)(0.12,-0.12){17}{\line(0,-1){0.12}}
\multiput(60.67,32.00)(-0.12,0.12){17}{\line(0,1){0.12}}
\multiput(64.67,28.00)(0.12,-0.12){17}{\line(0,-1){0.12}}
\put(18.00,51.00){\makebox(0,0)[cc]{$a$}}
\put(60.67,41.00){\framebox(4.00,5.00)[cc]{}}
\put(68.67,43.00){\makebox(0,0)[cc]{$\varphi$}}
\put(58.67,43.00){\makebox(0,0)[cc]{$P$}}
\put(70.67,33.00){\makebox(0,0)[cc]{$d$}}
\put(62.67,13.33){\oval(8.67,8.00)[b]}
\put(70.00,9.00){\makebox(0,0)[cc]{$D_2$}}
\put(65.33,20.33){\makebox(0,0)[cc]{$e$}}
\put(43.00,61.00){\makebox(0,0)[cc]{$b$}}
\put(62.00,61.00){\makebox(0,0)[cc]{$M$}}
\put(27.33,21.00){\makebox(0,0)[cc]{$M$}}
\end{picture}
\end{center}
\caption{Mach-Zehnder interferometer.
A single quantum (photon, neutron, electron {\it etc}) is emitted in $L$
and meets a lossless beam splitter (half-silvered mirror) $S_1$, after
which its wave function
is in a coherent superposition of $  b $ and $  c $. In beam
path $b$ a phase shifter shifts the phase of state $  b $ by
$\varphi$. The two beams are then recombined at a second lossless
beam splitter (half-silvered
mirror) $S_2$. The quant is detected at either $D_1$ or $D_2$,
corresponding to the states $d $ and $ e $, respectively.
 \label{f:m-z}}
\end{figure}
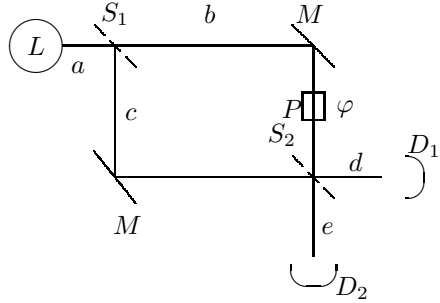

In what follows, a lossless {\em Mach-Zehnder} interferometer drawn in
Fig.
\ref{f:m-z} is discussed.
The computation proceeds by successive substitution (transition) of
states; i.e.,
\begin{eqnarray}
S_1:\; a  &\rightarrow& ( b  +i c
)/\sqrt{2}\quad , \\
P:\; b  &\rightarrow&  b e^{i \varphi
}\quad ,\\
S_2:\; b  &\rightarrow& ( e  + i
d )/\sqrt{2}\quad ,\\
S_2:\; c  &\rightarrow& ( d  + i
e )/\sqrt{2}\quad .
\end{eqnarray}
The resulting transition is
\begin{equation}
  a  \rightarrow \psi =i\left( {e^{i\varphi} +1\over
2}\right)
d  +
\left( {e^{i\varphi} -1\over 2}\right)
e  \quad .
\label{e:mz}
\end{equation}
Assume that $\varphi =0$, i.e., there is no phase shift at all.
Then, equation (\ref{e:mz}) reduces to
$ a  \rightarrow i d $, and the emitted quant is detected
only by $D_1$.
Assume that $\varphi =\pi $.
Then, equation (\ref{e:mz}) reduces to
$ a  \rightarrow -  e  $, and the emitted quant is detected
only by $D_2$.
If one varies the phase shift $\varphi$, one obtains the following
detection probabilities:
\begin{equation}
P_{D_1}(\varphi )=\vert ( d, \psi ) \vert^2=\cos^2({\varphi
\over 2})
\quad ,
\quad
P_{D_2}(\varphi )=\vert ( e, \psi ) \vert^2=\sin^2({\varphi
\over 2})
\quad .
\end{equation}

For some ``mindboggling'' features of Mach-Zehnder interferometry,
see \cite{benn:94}.

\section{Universal 2-port quantum gate}
\label{a:u(2)}

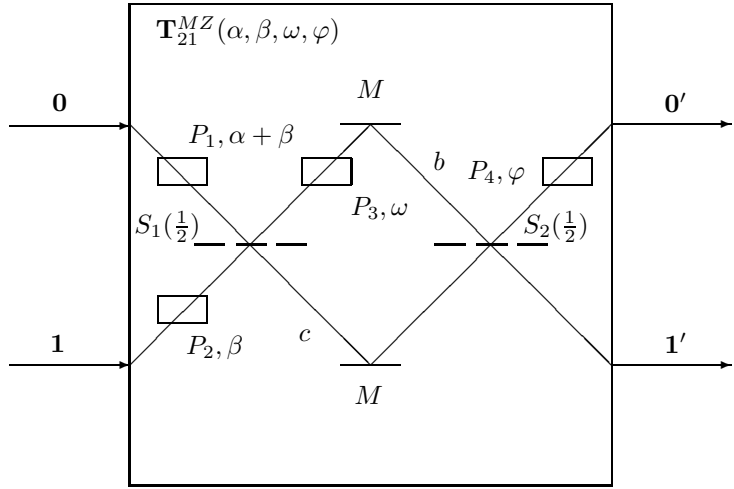
\begin{figure}
\begin{center}
\unitlength=0.80mm
\special{em:linewidth 0.4pt}
\linethickness{0.4pt}
\begin{picture}(120.00,200.00)
\put(20.00,120.00){\framebox(80.00,80.00)[cc]{}}
\put(57.67,160.00){\line(1,0){5.00}}
\put(64.33,160.00){\line(1,0){5.00}}
\put(50.67,160.00){\line(1,0){5.00}}
\put(78.67,170.00){\framebox(8.00,4.33)[cc]{}}
\put(82.67,178.00){\makebox(0,0)[cc]{$P_3,\varphi$}}
\put(73.33,160.00){\makebox(0,0)[lc]{$S(\omega )$}}
\put(8.33,183.67){\makebox(0,0)[cc]{${\bf 0}$}}
\put(110.67,183.67){\makebox(0,0)[cc]{${\bf 0}'$}}
\put(110.67,143.67){\makebox(0,0)[cc]{${\bf 1}'$}}
\put(8.00,143.67){\makebox(0,0)[cc]{${\bf 1}$}}
\put(24.33,195.67){\makebox(0,0)[lc]{${\bf T}_{21}^{bs}(\omega ,\alpha ,\beta
,\varphi )$}}
\put(0.00,179.67){\vector(1,0){20.00}}
\put(0.00,140.00){\vector(1,0){20.00}}
\put(100.00,180.00){\vector(1,0){20.00}}
\put(100.00,140.00){\vector(1,0){20.00}}
\put(20.00,14.67){\framebox(80.00,80.00)[cc]{}}
\put(20.00,34.67){\line(1,1){40.00}}
\put(60.00,74.67){\line(1,-1){40.00}}
\put(20.00,74.67){\line(1,-1){40.00}}
\put(60.00,34.67){\line(1,1){40.00}}
\put(55.00,74.67){\line(1,0){10.00}}
\put(55.00,34.67){\line(1,0){10.00}}
\put(37.67,54.67){\line(1,0){5.00}}
\put(44.33,54.67){\line(1,0){5.00}}
\put(30.67,54.67){\line(1,0){5.00}}
\put(77.67,54.67){\line(1,0){5.00}}
\put(84.33,54.67){\line(1,0){5.00}}
\put(70.67,54.67){\line(1,0){5.00}}
\put(88.67,64.67){\framebox(8.00,4.33)[cc]{}}
\put(85.67,66.67){\makebox(0,0)[rc]{$P_4,\varphi$}}
\put(60.00,80.67){\makebox(0,0)[cc]{$M$}}
\put(59.67,29.67){\makebox(0,0)[cc]{$M$}}
\put(31.67,57.67){\makebox(0,0)[rc]{$S_1({1\over 2})$}}
\put(85.33,57.67){\makebox(0,0)[lc]{$S_2({1\over 2})$}}
\put(8.33,78.34){\makebox(0,0)[cc]{${\bf 0}$}}
\put(110.67,78.34){\makebox(0,0)[cc]{${\bf 0}'$}}
\put(110.67,38.34){\makebox(0,0)[cc]{${\bf 1}'$}}
\put(8.00,38.34){\makebox(0,0)[cc]{${\bf 1}$}}
\put(49.00,39.67){\makebox(0,0)[cc]{$c$}}
\put(71.33,68.67){\makebox(0,0)[cc]{$b$}}
\put(24.33,90.34){\makebox(0,0)[lc]{${\bf T}_{21}^{MZ}(\alpha ,\beta
,\omega,\varphi )$}}
\put(0.00,74.34){\vector(1,0){20.00}}
\put(0.00,34.67){\vector(1,0){20.00}}
\put(100.00,74.67){\vector(1,0){20.00}}
\put(100.00,34.67){\vector(1,0){20.00}}
\put(48.67,64.67){\framebox(8.00,4.33)[cc]{}}
\put(56.67,60.67){\makebox(0,0)[lc]{$P_3,\omega$}}
\put(10.00,110.00){\makebox(0,0)[cc]{a)}}
\put(10.00,4.67){\makebox(0,0)[cc]{b)}}
\put(20.00,140.00){\line(2,1){80.00}}
\put(20.00,180.00){\line(2,-1){80.00}}
\put(32.67,170.00){\framebox(8.00,4.33)[cc]{}}
\put(36.67,178.00){\makebox(0,0)[cc]{$P_1,\alpha +\beta $}}
\put(24.67,64.67){\framebox(8.00,4.33)[cc]{}}
\put(29.67,72.67){\makebox(0,0)[lc]{$P_1,\alpha +\beta$}}
\put(24.67,41.67){\framebox(8.00,4.33)[cc]{}}
\put(29.34,37.67){\makebox(0,0)[lc]{$P_2,\beta$}}
\put(32.67,147.00){\framebox(8.00,4.33)[cc]{}}
\put(36.67,155.00){\makebox(0,0)[cc]{$P_2,\beta$}}
\end{picture}
\end{center}
\caption{Elementary quantum interference device.
An elementary quantum interference device can be realized by a
4-port interferometer with two input ports ${\bf 0} ,{\bf 1} $
and two
output ports
${\bf 0} ',{\bf 1} '$.
Any twodimensional unitary transformation can be realized by the
devices.
a) shows a realization
by a single beam
splitter $S(T)$
with variable transmission $t$
and three phase shifters $P_1,P_2,P_3$;
b) shows a realization with 50:50 beam
splitters $S_1({1\over 2}) $ and $S_2 ({1\over 2})$ and four phase
shifters
$P_1,P_2,P_3,P_4$.
 \label{f:qid}}
\end{figure}
The
elementary quantum interference device ${\bf T}_{21}^{bs}$ depicted in
Fig.
(\ref{f:qid}.a)
is just a beam splitter followed by a phase shifter in one of the output
ports. According to the ``toolbox'' rules of appendix
\ref{a:toolbox},
 the process can
be quantum mechanically described by\footnote{
Alternatively, the action of a lossless beam splitter may be
described by the matrix $
\left(
\begin{array}{cc}
T(\omega ) &i \, R( \omega )
\\
i\, R(\omega ) &T( \omega )
 \end{array}
\right)
=
\left(
\begin{array}{cc}
\cos \omega  &i \, \sin \omega
\\
i\, \sin \omega  &\cos \omega
 \end{array}
\right)
$.
A phase shifter in twodimensional Hilbert space is represented by
either
$
\left(
\begin{array}{cc}
e^{i\varphi }&0
\\
0&1
 \end{array}
\right)
$
or
$
\left(
\begin{array}{cc}
1&0
\\
0&e^{i\varphi }
 \end{array}
\right)
$.
 The action of the entire device consisting of such elements is
calculated by multiplying the matrices in reverse order in which the
quanta pass these elements \cite{yurke-86,teich:90}.
}
\begin{eqnarray}
P_1:\; {\bf 0}  &\rightarrow&  {\bf 0} e^{i
\alpha +\beta}
\quad , \\
P_2:\; {\bf 1}  &\rightarrow&  {\bf 1}
e^{i \beta}
\quad , \\
S:\; {\bf 0}
&\rightarrow& T\, {\bf 1}'  +iR\, {\bf 0}'
\quad , \\
S:\; {\bf 1}  &\rightarrow& T\, {\bf 0}'  +iR\,
{\bf 1}'
\quad , \\
P_3:\; {\bf 0}'  &\rightarrow&  {\bf 0}' e^{i
\varphi
}\quad .
\end{eqnarray}
If
$ {\bf 0}  \equiv    {\bf 0}' \equiv
\left(
\begin{array}{c}
1 \\
0
 \end{array}
\right)
$
and
$ {\bf 1} \equiv    {\bf 1}' \equiv
\left(
\begin{array}{c}
0 \\
1
 \end{array}
\right)
$ and $R(\omega )=\sin \omega $, $T(\omega )=\cos \omega$, then
the corresponding unitary evolution matrix
which transforms any coherent superposition of $ {\bf 0} $
and $ {\bf 1} $
into a superposition of
$ {\bf 0}' $
and
$ {\bf 1}' $
 is given by
\begin{eqnarray}
{\bf T}_{21}^{bs} (\omega ,\alpha ,\beta ,\varphi )&=&
\left[
e^{i\, \beta }\,
\left(
\begin{array}{cc}
i \, e^{i(\alpha +\varphi)} \,  \sin \omega  &e^{i\alpha }\,
  \cos \omega
\\ e^{i\varphi }\, \cos \omega  & i\,\sin \omega
 \end{array}
\right)
\right]^{-1}
\nonumber \\
&=&
e^{-i\, \beta }\,
\left(
\begin{array}{cc}
-i \, e^{-i(\alpha +\varphi)} \,  \sin \omega  &e^{-i\varphi }\,
  \cos \omega
\\ e^{-i\alpha }\, \cos \omega  & -i\,\sin \omega
 \end{array}
\right)
 \quad .
\label{e:quid1}
\end{eqnarray}

The
elementary quantum interference device ${\bf T}_{21}^{MZ}$ depicted in
Fig.
(\ref{f:qid}.b)
is a (rotated) Mach-Zehnder interferometer with {\em two}
input and output ports and three phase shifters.
According to the ``toolbox'' rules, the process can
be quantum mechanically described by
\begin{eqnarray}
P_1:\; {\bf 0}  &\rightarrow&  {\bf 0} e^{i
\alpha +\beta}
\quad , \\
P_2:\; {\bf 1}  &\rightarrow&  {\bf 1} e^{i
\beta}
\quad , \\
S_1:\; {\bf 1}  &\rightarrow& ( b  +i\,
c )/\sqrt{2}
\quad , \\
S_1:\; {\bf 0}  &\rightarrow& ( c  +i\,
b )/\sqrt{2}
\quad , \\
P_3:\; c  &\rightarrow&  c e^{i \omega
}\quad ,\\
S_2:\; b  &\rightarrow& ( {\bf 1}'  + i\,
{\bf 0}' )/\sqrt{2}\quad ,\\
S_2:\; c  &\rightarrow& ( {\bf 0}'  + i\,
{\bf 1}' )/\sqrt{2}\quad ,\\
P_4:\; {\bf 0}'  &\rightarrow&  {\bf 0}' e^{i
\varphi
}\quad .
\end{eqnarray}
When again
$ {\bf 0} \equiv   {\bf 0}' \equiv
\left(
\begin{array}{c}
1 \\
0
 \end{array}
\right)
$
and
$ {\bf 1} \equiv  {\bf 1}' \equiv
\left(
\begin{array}{c}
0 \\
1
 \end{array}
\right)
$, then
the corresponding unitary evolution matrix
which transforms any coherent superposition of $ {\bf 0} $
and $ {\bf 1} $
into a superposition of
$ {\bf 0}' $
and
$ {\bf 1}' $
 is given by
\begin{equation}
{\bf T}_{21}^{MZ} (\alpha ,\beta ,\omega ,\varphi )=
-i\,e^{-i(\beta +{\omega \over 2})}\;\left(
\begin{array}{cc}
-{e^{-i\,({\alpha +\varphi })}}\,\sin {{\omega }\over 2}
&
   {e^{-i\,{\varphi}}}\,\cos {{\omega }\over 2} \\
  e^{-i\,{\alpha }}\,\cos {{\omega }\over 2}&\sin {{\omega }\over
2}
 \end{array}
\right)
 \quad .
\label{e:quid2}
\end{equation}

The correspondence between
${\bf T}_{21}^{bs} (T(\omega ),\alpha ,\beta ,\varphi )$ with
${\bf T}_{21}^{MZ} (\alpha ',\beta ',\omega ',\varphi ')$ in equations
(\ref{e:quid1})
(\ref{e:quid2}) can be verified by comparing the elements of these
matrices.
The resulting four equations can be used to eliminate the four unknown
parameters
$\omega '=2\omega $,
$\beta '=\beta -\omega$,
$\alpha '=\alpha -\pi /2$,
$\beta '=\beta -\, \omega$ and
$\varphi '=\varphi -\pi /2$; i.e.,
\begin{equation}
{\bf T}_{21}^{bs} (\omega ,\alpha ,\beta ,\varphi ) =
{\bf T}_{21}^{MZ} (\alpha -{\pi \over 2}, \beta
-\omega ,2\omega
,\varphi
-{\pi \over 2})
\quad .
\end{equation}

Both elementary quantum interference devices are {\em universal} in the
sense that
{\em every} unitary quantum
evolution operator in twodimensional Hilbert space can be brought into a
one-to-one correspondence to  ${\bf T}^{bs}_{21}$ and
${\bf T}^{MZ}_{21}$; with corresponding values of
$T,\alpha ,\beta ,\varphi$ or
$\alpha ,\omega ,\beta ,\varphi $.
This can be easily seen by a similar calculation as before; i.e., by
comparing
equations
(\ref{e:quid1})
(\ref{e:quid2}) with the ``canonical''  form of a unitary matrix, which
is the product of a $U(1)=e^{-i\,\beta}$ and
 of the unimodular unitary
matrix $SU(2)$ \cite{murnaghan}
\begin{equation}
{\bf T} (\omega ,\alpha ,\varphi )=
\left(
\begin{array}{cc}
{e^{i\,\alpha }}\,\cos \omega
&
{-e^{-i\,\varphi }}\,\sin \omega
\\
{e^{i\,\varphi }}\,\sin \omega
&
{e^{-i\,\alpha }}\,\cos \omega
 \end{array}
\right)
 \quad ,
\end{equation}
where $-\pi \le \beta ,\omega \le \pi$,
$-\, {\pi \over 2} \le  \alpha ,\varphi \le {\pi \over 2}$.
Let
\begin{equation}
{\bf T} (\omega ,\alpha ,\beta ,\varphi )=
e^{-i\,\beta}{\bf T} (\omega ,\alpha ,\varphi )
\quad .
\end{equation}
A proper identification of the parameters
$\alpha ,\beta ,\omega ,\varphi $ yields
\begin{equation}
{\bf T} (\omega ,\alpha ,\beta ,\varphi )=
{\bf T}_{21}^{bs} (\omega -{\pi \over 2} ,-\alpha -\varphi -{\pi \over
2},
\beta + \alpha  +{\pi  \over 2} ,\varphi -\alpha +{\pi \over 2}
)
\quad .
\end{equation}

Let us examine the realization of a few primitive logical ``gates''
corresponding to (unitary) unary operations on qbits.
The ``identity'' element ${\Bbb I}$ is defined by
$ {\bf 0}  \rightarrow   {\bf 0} $,
$ {\bf 1}  \rightarrow   {\bf 1} $ and can be realized by
\begin{equation}
{\Bbb I} =
T^{bs}_{21}(-{\pi \over 2},-{\pi \over 2},{\pi \over 2},{\pi \over 2})=
T^{MZ}_{21}(-\pi ,\pi ,-\pi ,0)
=
\left(
\begin{array}{cc}
1&0
\\
0&1
 \end{array}
\right)
\quad .
\end{equation}

The ``${\tt not}$'' element is defined by
$ {\bf 0}  \rightarrow   {\bf 1} $,
$ {\bf 1}  \rightarrow   {\bf 0} $ and can be realized by
\begin{equation}
{\tt not} =
T^{bs}_{21}(0,0,0,0)=
T^{MZ}_{21}(-\,{\pi \over 2},0,0,-\,{\pi \over 2})
=
\left(
\begin{array}{cc}
0&1
\\
1&0
 \end{array}
\right)
\quad .
\end{equation}

The next element, ``$\sqrt{{\tt not}}$'' is a truly quantum
mechanical; i.e., nonclassical, one, since it converts a classical bit
into
a coherent superposition of $ {\bf 0} $ and $ {\bf 1} $.
$\sqrt{{\tt not}}$ is defined by
$ {\bf 0}  \rightarrow   {\bf 0}  +  {\bf 1} $,
$ {\bf 1}  \rightarrow  - {\bf 0}  +  {\bf 1} $ and can
be realized by
\begin{equation}
\sqrt{{\tt not}} =
T^{bs}_{21}(-{\pi \over 4},-{\pi \over 2},
{\pi \over 2},
{\pi \over 2})=
T^{MZ}_{21}(-\pi , {3\pi \over 4} ,-{\pi \over 2},0 )=
{1 \over \sqrt{2}}
\left(
\begin{array}{cc}
1&-1
\\
1&1
 \end{array}
\right)
\quad .
\end{equation}
Note that $\sqrt{{\tt not}}\cdot \sqrt{{\tt not}} = {\tt not}\cdot
{\rm diag}(1,-1)={\tt not}\, ({\rm mod }\, 1)$.
The relative phases in the output ports showing up in ${\rm
diag}(1,-1)$ can be avoided by defining
\begin{equation}
\sqrt{{\tt not}}' =
T^{bs}_{21}(-\,{\pi \over 4},0,
{\pi \over 4},
0)=
T^{MZ}_{21}(-\,{\pi \over 2} , {\pi \over 2} ,-\,{\pi \over 2},
-\,{\pi \over 2}
 )=
{1 \over 2}
\left(
\begin{array}{cc}
1+i&1-i
\\
1-i&1+i
 \end{array}
\right)
\quad .
\end{equation}
With this definition,
$
\sqrt{{\tt not}}'
\sqrt{{\tt not}}' = {\tt not}$.

It is very important that the elementary quantum
interference device realizes an arbitrary
quantum time evolution  of a twodimensional system.
The  performance of the quantum interference device is determined by
four parameters, corresponding to the phases
$\alpha ,\beta ,\varphi, \omega$.

\newpage
 \tableofcontents

\end{document}